\documentclass[a4paper,12pt]{article}
\usepackage{jheppub}
\usepackage{amssymb}
\usepackage{amsmath,amsfonts}
\usepackage{slashed}
\usepackage{amsthm}
\usepackage{graphicx}
\usepackage{hyperref}
\usepackage{rotating}
\usepackage{lscape}
\usepackage{pdflscape}
\usepackage{mathrsfs}
\usepackage{upgreek}
\usepackage{bm}
\usepackage{bbm}

\newcommand\bw{\begin{widetext}}
\newcommand\ew{\end{widetext}}
\newcommand{\be}{\begin{equation}}
\newcommand{\ee}{\end{equation}}
\newcommand{\beqa}{\begin{eqnarray}}
\newcommand{\eeqa}{\end{eqnarray}}

\newcommand\s{\sigma}
\renewcommand\a{\alpha}
\renewcommand\b{\beta}
\newcommand\g{\gamma}
\renewcommand\l{\lambda}

\newcommand\vk{\varkappa}

\def\e{{\rm e}}
\def\d{\partial}
\newcommand{\bseq}{\begin{subequations}}
\newcommand{\eseq}{\end{subequations}}

\renewcommand{\Im}{\mathop{\rm Im}\nolimits}
\renewcommand{\Re}{\mathop{\rm Re}\nolimits}

\newcommand{\D}{\mathcal D}

\newcommand{\DD}{\Delta}

\allowdisplaybreaks

\begin{document}
\begin{flushright}
CERN-TH-2016-219 \\
INR-TH-2016-038
\end{flushright}
\vspace{-1.5cm}

\title{Linearized supergravity with a dynamical preferred frame}
\author[a]{Arthur Marakulin}
\author[a,b,c,d,e]{and Sergey Sibiryakov}
\affiliation[a]{Institute for Nuclear Research of the
Russian Academy of Sciences, \\60th October Anniversary Prospect, 7a, 117312
Moscow, Russia}
\affiliation[b]{Department of Physics \& Astronomy, McMaster
  University,\\ Hamilton, Ontario, L8S 4M1, Canada}
\affiliation[c]{Perimeter Institute for Theoretical Physics, Waterloo,
  Ontario, N2L 2Y5, Canada}
\affiliation[d]{Theoretical Physics Department,
CERN, CH-1211 Gen\`eve 23, Switzerland}
\affiliation[e]{FSB/IPHYS/LPPC, \'Ecole Polytechnique
F\'ed\'erale de Lausanne,\\
CH-1015, Lausanne, Switzerland}
\emailAdd{marakulin@physics.msu.ru}
\emailAdd{ssibiryakov@perimeterinstitute.ca}

\abstract{We study supersymmetric extension of the Einstein-aether gravitational
model where local Lorentz invariance is broken down to the
subgroup of spatial rotations by a vacuum expectation value of a
timelike vector field called aether. Embedding aether into
a chiral vector superfield, we construct the most general
action which describes dynamics of linear perturbations around the
Lorentz-violating vacuum and is invariant under the linearized
supergravity transformations. The analysis is performed both in the
off-shell non-minimal superfield formulation of supergravity and in
the ``on-shell'' approach invoking only physical component fields. The
resulting model contains a single free coupling, in addition to the
standard supergravity parameters. The spectrum of physical excitations
features an enhanced on-shell gravity multiplet comprising four states
with helicities 2, 3/2, 3/2 and 1 propagating with superluminal
velocity. The remaining excitations propagate with the speed of
light. We outline the observational
constraints on the model following from its low-energy phenomenology.
}

\maketitle

\section{Introduction and summary}

The possibility to modify the laws of gravity has been the subject of
an intensive theoretical research during recent decades
\cite{Gregory:2000jc,
Dvali:2000hr,ArkaniHamed:2003uy,Dubovsky:2004sg,Rubakov:2008nh,
Nicolis:2008in,Deffayet:2010qz,deRham:2010kj}, see
\cite{Clifton:2011jh} for review.
This study has several motivations. First, it aims at solving the
problems faced by the Einstein's theory of general relativity (GR) at
very short distances, where it loses the predictive power because of
non-renormalizability, as well as at very long --- cosmological ---
scales where the standard paradigm leads to the cosmological constant
problem. Second, phenomenological models of modified gravity can be
used as proxies in the analysis of experimental data to put
constraints on deviations from GR at various scales within a
consistent framework. The third motivation is a deeper theoretical
understanding of the principles underlying GR and the consequences
implied by relaxing or replacing some of these principles.

An interesting class of modified gravity models involves violation of
the local Lorentz invariance. The possibility of such violation is
often attributed to the effects of quantum gravity, see
\cite{Mattingly:2005re,Liberati:2013xla} and references therein. In
particular, it has been suggested by P.~Ho\v rava \cite{Horava:2009uw}
that the quantum theory of gravity can be rendered perturbatively
renormalizable by abandoning Lorentz invariance as a fundamental
symmetry at high energies. The rigorous proof of renormalizability in a
version of this proposal has been given  in
\cite{Barvinsky:2015kil,Barvinsky:2017zlx}. In this framework some amount of Lorentz
symmetry breaking persists at all scales and at low energies the
theory reduces to GR coupled to a scalar field with non-zero timelike
gradient describing a preferred foliation of the
spacetime~\cite{Blas:2009yd,Blas:2009qj}.

Ho\v rava gravity is closely related
\cite{Jacobson:2010mx,Jacobson:2013xta,Blas:2010hb} to the
so-called
Einstein-aether model
\cite{Jacobson:2000xp,Jacobson:2008aj} where
the effects of the dynamical preferred frame are encoded by
a vector field $u_m$ (``aether'') constrained to have unit norm,\footnote{We use Latin
  letters from the middle of the alphabet for spacetime tensor
  indices; Latin letters from the beginning of the alphabet will be
  used for indices in the local Lorentz frame and Greek letters will
  be used for spinor indices. The signature of the metric is
  $(-,+,+,+)$; the Minkowski metric will be denoted by $\upeta_{mn}$.}
\begin{equation}
\label{unorm}
u_m u^m = -1\;.
\end{equation}
In the formal language, this vector belongs to the coset
$SO(3,1)/SO(3)$ of the Lorentz group over the group of spatial
rotations around the direction of $u_m$ that remain unbroken. This
construction is similar to the sigma-model description of
non-linearly realized simmetries in particle physics.
The most general action for the aether interacting with gravity and
containing up to two derivatives reads,
\begin{equation}
\label{Sae}
S = \frac{1}{2\vk^2}\int d^4 x \sqrt{-g}\big[R-K^{mn}_{\phantom{mn}sr}
\nabla_m u^s\nabla_n u^r+\l(u_mu^m+1)\big]\;,
\end{equation}
where $\lambda$ is a Lagrange multiplier enforcing the constraint
(\ref{unorm}) and
\be
\label{K}
K^{mn}_{\phantom{mn}sr} \equiv
c_1g^{mn}g_{sr}+c_2\delta_s^m\delta_r^n+c_3\delta_r^m\delta_s^n
-c_4u^m u^ng_{sr}\;.
\ee
The theory contains four dimensionless parameters $c_i$,
$i=1,2,3,4$. When the constraint (\ref{unorm}) is solved explicitly
and the action is written in terms of independent components of $u_m$,
it contains non-linear derivative self-interactions of these
components. This restricts the domain of validity of the model to
energies
below $M_*\equiv\vk^{-1}\sqrt{c}$, where $c$ is the characteristic
value of the couplings $c_i$. At higher energies the model becomes
strongly coupled and requires an ultraviolet (UV) completion. By
analogy with sigma-models, the scale $M_*$ can be identified with the
scale of the Lorentz symmetry breaking,\footnote{It is worth stressing
  that, unlike the case of the usual spontaneous
  symmetry breaking,
  Lorentz invariance need not be restored above $M_*$. On the
  contrary, violation of Lorentz invariance can become increasingly
  more important at high energies, as it happens in Ho\v rava
  gravity.} the product $(\vk M_*)^2=c$
controlling the strength of Lorentz violating effects in gravity.
Phenomenology of this model has been extensively studied
resulting in constraints on the couplings $c_i$
\cite{Carroll:2004ai,Foster:2005dk,Jacobson:2008aj,Yagi:2013qpa,Yagi:2013ava,
Audren:2014hza},
see \cite{Blas:2014aca} for review.
It was also proposed to use Ho\v rava
gravity and Einstein-aether models for holographic description of
strongly coupled non-relativistic systems
\cite{Janiszewski:2012nb,Janiszewski:2012nf,Griffin:2012qx}.

It has long been envisaged that an important role at high energies
can be played by supersymmetry (SUSY). In particle physics SUSY is
usually considered as an extension of the Poincar\'e group. However,
as pointed out in \cite{GrootNibbelink:2004za,Bolokhov:2005cj}, the SUSY algebra
reduced by removing the boost generators closes on itself. In other
words, SUSY does not necessarily require Lorentz
invariance. Conversely, a general non-relativistic SUSY consisting of
space- and time-translations, spatial rotations and supercharges in
the spinor representation of $SO(3)$ is equivalent to the standard
SUSY algebra without boosts \cite{Pujolas:2011sk}. Remarkably, SUSY
enforces emergence of Lorentz symmetry at low energies in the Standard
Model, even if the high-energy theory is not Lorentz invariant
\cite{GrootNibbelink:2004za,Bolokhov:2005cj,Pujolas:2011sk}. This could explain the
exquisite precision with which Lorentz invariance is satisfied in
particle physics\footnote{Soft breaking of SUSY with superpartner
  masses parametrically below the scale of Lorentz violation
 preserves the suppression of
  Lorentz-violating effects in the Standard Model
  \cite{GrootNibbelink:2004za,Bolokhov:2005cj,Pujolas:2011sk}.}
\cite{Mattingly:2005re,Kostelecky:2008ts,Liberati:2013xla}.

It is natural to ask whether the local generalization of SUSY leading
to the theory of supergravity (SUGRA) is also compatible with the
existence of a preferred frame. Clearly, as in the case of ordinary
gravity, this frame must be dynamical. The first step
in answering this question was made in
Ref.~\cite{Pujolas:2011sk} which has constructed the supersymmetric
extension of the aether model in flat
spacetime. In the superspace formalism, the aether is
promoted to a chiral vector superfield $U^c$,
\be
\label{chir}
\bar D_{\dot\a}U^c=0\;,
\ee
where $\bar D_{\dot\a}$ is the superspace covariant
derivative.\footnote{We
  use the notations and conventions of \cite{WB} for the objects
  related to the spinor algebra and superspace geometry.}
The lowest component of $U^c$ is identified with the complexified
aether $u^c$. This choice of multiplet is motivated by the following
considerations. Imposing on the superfield 
a constraint similar to (\ref{unorm}),
\be
\label{Unorm}
U_cU^c=-1\;,
\ee
forces aether to develop a vacuum expectation value (VEV) that breaks
the Lorentz symmetry. 
As a consequence, the latter is realized
non-linearly on the perturbations around the vacuum. 
On the other
hand, since the aether is the lowest component of the superfield, its
VEV preserves SUSY which remains linearly
realized.\footnote{This is different from the
setup considered in the context of supersymmetric effective theory of
  inflation~\cite{Baumann:2011nk,Delacretaz:2016nhw}
where not only the Lorentz group, but also SUSY
  is realized non-linearly.} This is interesting from the
phenomenological perspective as it admits the mechanism of
Refs.~\cite{GrootNibbelink:2004za,Bolokhov:2005cj,Pujolas:2011sk} for
protection of Lorentz invariance in the matter sector. A different
embedding (i.e. not as the lowest component of a supermultiplet) would
break SUSY together with Lorentz invariance 
and hence lead to unsuppressed propagation of
Lorentz violation to the matter fields.  
In addition to the aether, the superfield $U^c$ describes its superpartner
--- ``aetherino''.

Upon
an eventual
soft SUSY breaking aetherino and the imaginary part of the aether
acquire masses, whereas the action for the real part reduces to the
flat-spacetime limit of (\ref{Sae}) with a special choice of the
couplings,\footnote{Soft SUSY
  breaking introduces corrections to these relations suppressed by the
SUSY breaking scale.}
\begin{equation}
\label{c234}
c_2 + c_3 = c_4=0\;.
\end{equation}
The coupling $c_1$ remains unrestricted.
The analysis in flat spacetime is insufficient to decide
whether SUSY constrains the parameters
$c_2$ and $c_3$
separately or just their sum. For the Minkowski metric the $c_2$ and
$c_3$ terms in the aether Lagrangian differ by a total derivative and
only the sum $c_2+c_3$ remains in the action after integration by parts.

In this paper we construct the interaction of the super-aether theory
of Ref.~\cite{Pujolas:2011sk} with linearized supergravity. 
The latter is invariant only under the linearized version of the local
Lorentz symmetry. Hence, we also need to linearize 
the super-aether field describing the
spontaneous breaking of this symmetry. In other words, we
expand the super-aether into a constant background $w^c$
and fluctuations $V^c$,
\begin{equation}
\label{Vdef}
U^c= w^c + V^c\;,
\end{equation}
and keep terms up to quadratic order in $V^c$ in the action.

The reason
to focus on the linearized theory is twofold. 
First, this restriction allows us
to bypass the complications associated to the construction of the
curved superspace \cite{WB,Girardi:1984eq,Gates:1983nr,buch} and 
use instead the transparent formalism of flat superspace. In this way
we classify all
possible supersymmetric terms in the Lagrangian and thereby analyze
the uniqueness of the action. 
We find that local SUSY highly constrains the aether
sector leaving only a single free parameter that is identified with
$c_1$ in the low-energy theory, whereas the parameters $c_2$ and $c_3$
are forced to vanish.
Second, this path directly leads us to the quadratic action for
perturbations in components, which we use to analyze the dispersion
relations of various physical modes.

We will follow two complementary approaches: the fully off-shell
superfield formalism and the ``on-shell'' approach where one works
only with physical component fields. In complete analogy with
linearized gravity that is invariant both under linearized
diffeomorphisms and the global Poincar\'e group, linearized
SUGRA prossesses two sets of supersymmetries: global and local
ones. The superfield formalism has the advantage of manifestly
implementing the global SUSY as translations in the superspace. On the
other hand, the local SUSY transformations are more complicated. They
are encoded in linearized super-diffeomorphisms acting as gauge
transformations on the superfields. The superspace gauge
group is, however, too large. Its partial gauge fixing down to the
physically relevant subgroup requires augmenting the auxiliary sector
with a compensator multiplet.

The chirality constraint that we want to impose on the aether
superfield forces us to use the non-minimal off-shell formulation of
SUGRA \cite{Breitenlohner:1976nv,Breitenlohner:1977jn,Siegel:1978mj}.
To see this, recall the general form of the
anti-commutator of two spinor derivatives acting on a vector
superfield~\cite{WB},
\begin{equation}
\label{DaDb}
\{\bar\D_{\dot\alpha},
\bar\D_{\dot\beta}\}U^c=-T_{\dot\a\dot\b}^{\phantom{\dot\a\dot\b}b}
\D_bU^c
-T_{\dot\a\dot\b}^{\phantom{\dot\a\dot\b}\g}
\D_\g U^c
-T_{\dot\a\dot\b}^{\phantom{\dot\a\dot\b}\dot\g}
\bar\D_{\dot\g} U^c
+U^bR_{\dot\a\dot\b b}^{\phantom{\dot\a\dot\b b}c}\;,
\end{equation}
where $T_{AB}^{\phantom{AB}C}$ and $R_{ABC}^{\phantom{ABC}D}$ are
respectively torsion and curvature in the superspace.\footnote{Capital
  Latin letters $A,B,\ldots$ are used for the general superspace
  indices.} In the minimal SUGRA all components of the torsion
appearing in (\ref{DaDb}) vanish, whereas $R_{\dot\a\dot\b bc}$ is in
general non-zero~\cite{WB}. This implies that
the chirality constraint cannot be imposed in a covariant way as it is
incompatible with
(\ref{DaDb}). On the other hand, in the non-minimal formulation
the supercovariant derivatives can
be chosen such that \cite{Brown:1979xt,Girardi:1984eq}
\be
\label{nonminR}
T_{\dot\a\dot\b}^{\phantom{\dot\a\dot\b}b}
=T_{\dot\a\dot\b}^{\phantom{\dot\a\dot\b}\g}
=R_{\dot\a\dot\b b}^{\phantom{\dot\a\dot\b b}c}=0
\ee
and therefore the covariant chirality constraint
\be
\label{covchir}
\bar\D_{\dot\a}U^c=0
\ee
is consistent with (\ref{DaDb}).
We classify all terms in the superspace Lagrangian that can be
constructed from the linearized SUGRA and aether superfields and then
fix the coefficients in front of them by imposing the invariance under
super-diffeos.

In the ``on-shell'' approach the situation is in a sense reverse. Here the local
gauge invariance is realized as simple gradient transformations of the
metric and gravitino. The construction of the aether--supergravity
interaction amounts to the classification of the possible
aether energy-momentum tensors (EMTs) and supersymmetry
currents coupled respectively to the metric perturbations
and gravitino. The
interaction, however, deforms global SUSY transformations and the
invariance of the Lagrangian with respect to them must be checked
explicitly.
We work out the on-shell SUSY and perform this check perturbatively
up to second order
in the strength of the aether--gravity coupling.

The paper is organized as follows. In Sec.~\ref{sec:2} we review the
formalism of linearized non-minimal SUGRA.
While most of this material is standard, we present new derivations of
several relations that are used in the rest of the paper. In particular, we
work out in detail the gauge fixing in non-minimal linearized SUGRA
and give explicit expressions for the superspace connection in terms
of the gravity and compensator superfields.
Supplementary material for this section is contained in
Appendices~\ref{app:globalSUSY} and \ref{app:torsion}.

In Sec.~\ref{sec:3} we
introduce breaking of Lorentz invariance and
the linearized aether superfield. We classify all inequivalent
terms in the quadratic Lagrangian, derive their variations
under super-diffeomorphisms
and find the most general invariant superfield action.
Appendix~\ref{app:useful} contains technical details of this derivation.

In Sec.~\ref{sec:bosonic} we derive the bosonic part of the Lagrangian
in components. We first present the full off-shell action and then
integrate out the auxiliary fields perturbatively in the aether
coupling.

In Sec.~\ref{sec:fermionic} (complemented with
Appendices~\ref{app:supercurrent}, \ref{app:omega}) we switch to the ``on-shell''
formalism and work out the most general form of the linearized aether
EMT and supercurrent. This provides us with an alternative proof of
the uniqueness of
the aether--supergravity coupling. We reconstruct the fermionic part
of the Lagrangian and derive the on-shell SUSY transformations.

In
Sec.~\ref{sec:physical} we analyze some physical implications of the
model. Notably, we find that the velocity of gravitons necessarily
exceeds the speed of light,\footnote{Superluminal propagation of signals does not lead
    to any inconsistencies in theories with a preferred reference frame. In
    particular, absence of Lorentz invariance at the fundamental level
    prevents from creating closed timelike curves, see e.g. discussion
    in \cite{Jacobson:2008aj,Foster:2005dk}.}
thereby manifesting violation of the
Lorentz invariance in the low-energy theory.
This is in contrast with the Standard Model matter and gauge fields
which are protected by SUSY from Lorentz symmetry violation at low
energies~\cite{GrootNibbelink:2004za,Bolokhov:2005cj}. The excess of
graviton velocity over unity is interesting from the viewpoint of SUSY
representations as it leads to an enhancement of the graviton multiplet
by additional states with helicities 3/2 and 1. The dispersion
relations for the correspoding modes are derived in
Appendix~\ref{app:disph1} and they are shown to match that of the
graviton.
We also briefly
discuss the phenomenological constraints on the model.

Sec.~\ref{sec:conclusions} is devoted to conclusions.

\section{Linearized non-minimal supergravity}
\label{sec:2}

\subsection{Field content and gauge fixing}
\label{sec:fieldcontent}

We follow \cite{Gates:1983nr,buch}. The basic ingredient of the linearized
SUGRA is a real vector superfield $H_m$ transforming
as\footnote{Throughout the text $D_\a$, $\bar D_{\dot\a}$ denote the
  covariant derivatives of the flat superspace preserving
the global SUSY. They should not be
  confused with the derivatives covariant under local SUSY
  transformations that are denoted by $\D_\a$, $\bar\D_{\dot\a}$.}
\begin{equation}
\label{Htrans}
\delta_L H_{\alpha {\dot \alpha}} =
{\bar D}_{\dot \alpha} L_\alpha - D_\alpha {\bar L}_{\dot \alpha}
\end{equation}
under the linearized super-diffeomorphisms parameterized by the spinor
superfield $L_\a$. To understand the physical content of $H_m$ let us
decompose it in components~\cite{Linch:2002wg},
\bseq
\label{Hcomp}
\begin{align}
&c_m=H_m\big|\;,
&&\chi_{\a\b\dot\b}=D_\a H_{\b\dot\b}\big|\;,
&&a_m=-\frac{1}{4}D^2H_m\big|\;,\\
&e_{mn}=-\Delta_mH_n\big|\;,
&&\psi^m_\a=\frac{i}{16}\bar\s^{m\dot\b\b}\bar D^2D_\b
H_{\a\dot\b}\big|\;,
&&d_m=\frac{1}{32}\{D^2,\bar D^2\}H_m\big|\;,
\end{align}
\eseq
where the vertical line denotes evaluation at zero fermionic
coordinates $\theta=\bar\theta=0$, and
\be
\label{deltadef}
\Delta_m H_n\equiv\frac{1}{4}\bar\s_m^{\dot\a\a}[\bar D_{\dot\a},D_\a]H_n\;.
\ee
We see that the multiplet contains a spin-2 field $e_{mn}$ that is
identified with the perturbation of the tetrad, as well as the
spin-3/2 field $\psi^m_\a$ describing gravitino. Introducing
also the components of the gauge parameter $L_\a$,
\bseq
\label{Lcomp}
\begin{align}
&\xi_m\equiv\xi_m^R+i\xi_m^I=i\bar\s_m^{\dot\a\a}\bar D_{\dot\a}L_\a\big|\;,
~~~~\varepsilon_{\a}=-\frac{1}{4}\bar D^2L_\a\big|\;,
~~~~\mu^m_\a=i\bar\s^{m\dot\b\b}D_\a\bar D_{\dot\b}L_\b\big|\;,\\
&\l_\a^{\phantom\a \b}=-\frac{1}{4}D_\a\bar D^2L^\b\big|\;,
~~~~~\kappa^m=-\frac{i}{4}\s^{m}_{\a\dot\a}D^2\bar D^{\dot\a}L^\a\big|\;,
~~~~~\rho_\a=\frac{1}{16}D^2\bar D^2L_\a\big|\;,
\end{align}
\eseq
one obtains from (\ref{Htrans}) the following transformation laws:
\bseq
\label{comptrans}
\begin{align}
\label{comptrans1}
&\delta_L c_m=-\xi_m^I\;,~~~~~~
\delta_L\chi_{\a\b\dot\b}=\frac{i}{2}\s_{m\b\dot\b}\mu^m_\a+2\epsilon_{\a\b}
\bar\varepsilon_{\dot\b}\;,~~~~~~
\delta_L a_m=\frac{i}{2}\kappa_m\;,\\
\label{comptrans2}
&\delta_L e_{mn}=\d_m\xi_n^R+(\s_{mn})^{\a\b}\l_{\a\b}
-(\bar\s_{mn})^{\dot\a\dot\b}\bar\l_{\dot\a\dot\b}
+\frac{1}{2}\upeta_{mn}(\l^\a_{~\a}-\bar\l_{\dot\a}^{~\dot\a})\;,\\
\label{comptrans3}
&\delta_L\psi^m_\a=\d^m\varepsilon_\a-\frac{i}{2}\s^{m}_{\a\dot\b}\,\bar\rho^{\dot\b}\;,
~~~~~~\delta_L d_m=-\frac{1}{2}\Box\xi_m^I+
\Big[\frac{i}{4}(\sigma^n\bar\sigma_m)_{\a\g}\d_n\lambda^{\a\g}+\mathrm{h.c.}\Big]\;,
\end{align}
\eseq
where $\Box\equiv\upeta^{mn}\d_m\d_n$.
It follows from (\ref{comptrans1}) that the imaginary part of $\xi_m$
and the components $\mu^m_\a$, $\kappa_m$ can be chosen to impose the
Wess--Zumino gauge,
\be
\label{WZ}
c_m=\chi_{\a\b\dot\b}=a_m=0\;.
\ee
Note that this implies the relation between the gauge parameters
$\mu^m_\a$ and $\varepsilon_\a$,
\be
\label{mueps}
\mu^m_\a=2i\sigma^m_{\a\dot\b}\,\bar\varepsilon^{\dot\b}\;.
\ee
The remaining transformations contain infinitesimal diffeomorphisms
with the parameter $\xi_m^R$, local Lorentz transformations
parameterized by the symmetric part of $\l_{\a\b}$ and local SUSY
corresponding to $\varepsilon_\a$. The trace part $\l_{~\a}^\a$ and the
spinor $\rho_\a$ give rise to extra symmetries: Weyl invariance and
superconformal transformations. The latter symmetries are not
generally present
in SUGRA. They are removed by introducing a compensator.

In the minimal linearized SUGRA the compensator is chosen to be a
chiral scalar superfield. However, as discussed in the Introduction,
the minimal formulation does not admit a coupling to the super-aether
theory. The next-to-simplest choice of the compensator, which leads
to the non-minimal formulation, is a linear superfield $\Gamma$,
\be
\label{linGamma}
\bar D^2\Gamma=0\;,
\ee
with independent components,
\bseq
\label{compGamma}
\begin{align}
\label{lowcompGamma}
&\g=\Gamma\big|\;,
&&\bar\omega_{\dot\a}=\bar D_{\dot\a}\Gamma\big|\;,
&&\phi_\a=D_\a\Gamma\big|\;,\\
\label{uppercompGamma}
&B=-\frac{1}{4}D^2\Gamma\big|\;,
&&q_m\equiv q_m^R+iq_m^I=\Delta_m\Gamma\big|\;,
&&\bar\nu_{\dot\a}=\frac{1}{4}D^2\bar D_{\dot\a}\Gamma\big|\;.
\end{align}
\eseq
It transforms under the super-diffeomorphisms as
\begin{equation}
\label{Gammatrans}
\delta_L \Gamma = -\frac{n+1}{4(3n+1)}\bar D^2D^\a L_\a+
\frac{1}{4} {\bar D}^{\dot\alpha} {D}^2 {\bar L}_{\dot\alpha}\;,
\end{equation}
where $n\neq -\frac{1}{3},0$ is a real parameter enumerating different
versions of the non-minimal SUGRA.
The gauge transformations of the components read,
\bseq
\label{lowGtrans}
\begin{align}
&\delta_L\g=\frac{n+1}{3n+1}\partial_m\xi^m+\frac{n+1}{3n+1}\l^\a_{~\a}
+\bar\l^{\dot\a}_{~\dot\a}\;,~~~~~~~\delta_L\bar\omega_{\dot\a}=2\bar\rho_{\dot\a}\;,\\
&\delta_L\phi_\a=\frac{2(n+1)}{3n+1}\rho_\a+\frac{n+1}{3n+1}\d_m\mu^m_\a
-2i\sigma^m_{\a\dot\a}\d_m\bar\varepsilon^{\dot\a}\;,
~~~~~\delta_L B=\frac{n+1}{3n+1}\d_m\kappa^m\;,\\
&\delta_Lq_m=\frac{i(n+1)}{3n+1}\d_m\d_n\xi^n
+\frac{i(n+1)}{3n+1}\d_m\lambda^\a_{~\a}-2i(\bar\sigma_{mn})^{\dot\a\dot\b}
\d^n\bar\lambda_{\dot\a\dot\b}\;,
~~~\delta_L\bar\nu_{\dot\a}=-2\Box \bar\varepsilon_{\dot\a}\;.
\end{align}
\eseq
They suggest to fix away the gauge parameters $\lambda^\a_{~\a}$ and
$\rho_\a$ by imposing the conditions,
\bseq
\label{compfix}
\begin{align}
\label{weylfix}
&\g=\frac{n+1}{3n+1}e^m_{~\;m}\;,\\
\label{superconffix}
&\omega_{\a}+\phi_{\a}+\frac{4in}{3n+1}\sigma^m_{\a\dot\a}\,\bar\psi_m^{\dot\a}=0\;.
\end{align}
\eseq
Note that the coefficient in front of
the term with gravitino in Eq.~(\ref{superconffix})
has been chosen in such a way that the contributions
with $\bar\varepsilon^{\dot\a}$ cancel out from the transformation of the
l.h.s. when the relation (\ref{mueps}) is imposed.
The choice of the relative coefficients between $\omega_\a$ and
$\phi_\a$ will be discussed shortly.

Further, we can use the symmetric part of $\lambda_{\a\beta}$ to
render the tetrad symmetric,
\be
\label{sime}
e_{mn}=e_{nm}=\frac{1}{2}h_{mn}\;,
\ee
where $h_{mn}$ are the perturbations of the metric. To preserve this gauge,
the parameters $\lambda_{\a\b}$ must be related to the
linearized local translations as
\be
\label{lamxi}
(\sigma_{mn})^{\a\b}\lambda_{\a\b}-(\bar\sigma_{mn})^{\dot\a\dot\b}
\bar\lambda_{\dot\a\dot\b}=-\frac{1}{2}\d_m\xi_n^R+\frac{1}{2}\d_n\xi_m^R\;.
\ee
This completes the gauge fixing procedure that will be used when
considering the formulation of the theory in components.
It leaves only the linearized
diffeomorphisms and the local SUSY transformations, under which the
metric perturbations and garvitino transform in the standard way,
\be
\label{standlocal}
\delta_L h_{mn}=\d_m\xi_n^R+\d_n\xi_m^R~,~~~~~~~~
\delta_L\psi_\a^m=\d^m\varepsilon_\a\;.
\ee

In addition to the local super-diffeomorphisms, the linearized supergravity
is invariant under the global SUSY transformations, whose
coordinate-independent parameter will be denoted by $\zeta^\a$. This
symmetry is manifest in the superfield formalism. In
Appendix~\ref{app:globalSUSY} we work out the global SUSY
transformations of the component fields and discuss how they are
affected by
the gauge fixing. In particular, we show that the special choice of
coefficients in the gauge condition (\ref{superconffix}) brings
the SUSY transformation of the metric into the canonical form
in terms
of gravitino \cite{WB},\footnote{Note that our definition of gravitino differs
  by a factor $-1/2$ from that adopted in \cite{WB}.}
\be
\label{tildedGh}
\tilde\delta_G
h_{mn}=2i(\zeta\sigma_m\bar\psi_n+\zeta\sigma_n\bar\psi_m
+\bar\zeta\bar\sigma_m\psi_n+\bar\zeta\bar\sigma_n\psi_m)\;.
\ee

The action of the linearized non-minimal SUGRA has the form~\cite{buch},
\bseq
\label{NMact}
\be
S_{SG} =\int d^4x\int d^2\theta d^2\bar\theta \;\mathbb{L}_{SG}\;,
\ee
with the superspace Lagrangian,
\be
\begin{split}
&\mathbb{L}_{SG} = \frac{1}{\vk^2}
\bigg[\frac{1}{4} \Big((\partial_k H_m)^2 -
(\Delta_k H_m)^2\Big)
+\frac{n+1}{2n}(\d_mH^m)^2+\frac{n+1}{2}(\Delta_mH^m)^2\\
&-i\frac{3n\!+\!1}{2n}\partial_m H^m \left(\Gamma \!-\! \bar\Gamma\right)
+\frac{3n\!+\!1}{2} \Delta_m H^m \left(\Gamma\! +\! \bar\Gamma\right)
+\frac{9n^2\!-\!1}{8n}(\Gamma^2\! +\! \bar \Gamma^2)
+\frac{(3n\!+\!1)^2}{4n} \Gamma\bar\Gamma\bigg].
\end{split}
\ee
\eseq
It is straightforward to verify that it is invariant under the
transformations (\ref{Htrans}), (\ref{Gammatrans}). The above
expressions simplify considerably for the choice $n=-1$. However, we
are not going to restrict to this case as we want to study the most
general coupling of the super-aether to gravity.

Using the general formula
\be
\label{LsupL}
{\cal L}=\frac{1}{32}\{\bar D^2,D^2\}\,\mathbb{L}\big|\;
\ee
that relates the component Lagrangian ${\cal L}$ to that in superspace
one obtains the off-shell Lagrangian of non-minimal SUGRA in
  terms of the component fields. Its bosonic part reads,
\be
\label{LSG}
\begin{split}
{\cal L}_{SG}^{\rm bos}
=\frac{1}{2\vk^2}&\bigg\{\frac{1}{4}h_{km}\Box h_{km}
+\frac{n-1}{8n}\d_k h_{km}\d_lh_{lm}
-\frac{n+1}{4(3n+1)} h\Box h\\
&+\frac{3n+1}{2n}q^I_m\d_k h_{km}
-q_m^I\d_m h
-\frac{3n+1}{2n} q^I_m q^I_m\\
&-2(n-1) d_m d_m +2(3n+1)d_m q^R_m
-\frac{3(3n+1)}{2} q^R_m q^R_m
+\frac{(3n+1)^2}{2n} B\,\bar B\bigg\},
\end{split}
\ee
where $h\equiv h_m^m$.
After integrating out the auxiliary fields $q_m$, $d_m$ and $B$ it takes the
well-known form of the linearized Einstein--Hilbert action
\be
\label{LSGHG}
{\cal L}_{EH}=\frac{1}{2\vk^2}\left(\frac{1}{4}h_{km}\Box
  h^{km}+\frac{1}{2}\d^kh_{km}\d_lh^{lm}-\frac{1}{2}\d_kh^{km}\d_mh
+\frac{1}{4}\d_mh\d^mh\right)\;.
\ee
The fermionic part of the action upon elimination of the
auxiliary fields is also well-known and is given by the
Rarita--Schwinger Lagrangian,
\be
\label{Rarita}
{\cal L}_{RS}=\frac{4}{\vk^2}\,\epsilon^{klmn}\,\bar\psi_k\,\bar\sigma_l\,\d_m\psi_n\;,
\ee
where $\epsilon^{klmn}$ is the totally antisymmetric tensor, $\epsilon^{0123}=1$.

\subsection{Lorentz  transformations in superspace and
covariant derivatives}
\label{sec:2.2}

In non-linear superspace formulation of SUGRA
the most general transformation of a superfield
consists of super-diffeomorphisms and local Lorentz
rotations~\cite{WB}. The particular realization of the linearized
supergravity using the fields $H_m$, $\Gamma$ corresponds to a
(partial) gauge fixing of this symmetry linking the superspace
translations and local Lorentz rotations to the single spinor
superfield $L_\a$. Thus, the transformations of a scalar superfield
$\Psi$ and a superfield with a Lorentz index $\Psi^A$ read
respectively,
\[
\delta_L\Psi=l^M(L_\a)\,\d_M\Psi~,~~~~~~
\delta_L\Psi^A=l^M(L_\a)\,\d_M\Psi^A+\Psi^B\,M_{B}^{~\;A}(L_\a)\;,
\]
where capital letters denote general --- vector or spinor ---
indices. The only non-vanishing components of the matrix $M_B^{~\;A}$
are
$M_{\b}^{~\;\a}$, $M^{\dot\b}_{~\;\dot\a}$, $M_{b}^{~\;a}$ and
satisfy the structural relations of the $SL(2)$ algebra,
\be
\label{structrel}
M_{\a\b}=M_{\b\a}~,~~~~~ M_{\dot\a\dot\b}=-(M_{\a\b})^*~,~~~~~
M_{ab}=\frac{1}{2}\bar\s_a^{\dot\a\a}\bar\s_b^{\dot\b\b}
(\epsilon_{\dot\a\dot\b}M_{\a\b}-\epsilon_{\a\b}M_{\dot\a\dot\b})\;.
\ee
Our present goal is to work out the expressions for $M_{A}^{~\;B}$ in
terms of $L_\a$.

We start from the expression for the differential operator describing
local superspace translations given in
Ref.~\cite{Linch:2002wg},\footnote{We choose the
  representation of the super-diffeos that preserves real superfields.}
\be
\label{scaltrans}
\hat L\equiv l^M(L_\a)\,\d_M =-\frac{1}{4}(\bar D^2L^\a)D_\a
-\frac{1}{4}(D^2\bar L_{\dot\a})\bar D^{\dot\a}
+\frac{i}{2}(\bar D^{\dot\a}L^\a+D^\a\bar
L^{\dot\a})\,\d_{\a\dot\a}\;.
\ee
As discussed in \cite{Linch:2002wg}, this form is fixed by the
requirement that scalar chiral superfields should admit covariant
generalizations. Covariant derivatives of a scalar must transform as
spinors or vectors. In particular,
\be
\label{Dtrans1}
\delta_L({\cal D}_\a\Psi)=\hat{L}{\cal D}_\a\Psi
+({\cal D}^\b\Psi)\, M_{\b\a}\;,
\ee
On the other hand, the covariant derivatives are related to the
ordinary derivatives through the superspace vielbein $E_A^{~\;M}$,
\be
\label{covdergen}
{\cal D}_A\Psi=E_A^{~\;M}\,\d_M\Psi\;.
\ee
At linear order the vielbein can be written as
\be
\label{viellin}
E_A^{~\;M}=E_A^{(0)\,M}-I_A^{~\;B}\,E_B^{(0)\,M}\;,
\ee
where $E_A^{(0)\,M}$ is the vielbein of the flat superspace.
On general grounds, the tensor $I_A^{~\;B}$ must be linear in the
SUGRA fields\footnote{At the
  linearized level we do not distinguish the spacetime and Lorentz
  indices whenever it does not lead to confusion.}
$H_b$,
$\Gamma$, with the precise expressions yet to be determined.
Substituting Eqs.~(\ref{covdergen}), (\ref{viellin}) into the
l.h.s. of Eq.~(\ref{Dtrans1}) and using $E_A^{(0)\,M}\d_M\Psi=D_A\Psi$
we obtain,
\[
\delta_L({\cal D}_\a\Psi)=\delta_L(D_\a\Psi-I_\a^{~B}D_B\Psi)
=D_\a\hat L\Psi-(\delta_L I_\a^{~B})D_B\Psi\;,
\]
where in the second equality we neglected terms of the form
$I_\a^{~B}D_B\hat L\Psi$ as they are quadratic in the deviations
from the flat superspace geometry. Equating the coefficients in
front of
different derivatives $D_A\Psi$ in Eq.~(\ref{Dtrans1})
we obtain a system of equations for
$M_{\a\b}$ and the variations of the vielbein components
$I_\a^{~\;\b}$, $I_{\a\dot\b}$, $I_\a^{~\;b}$. To solve this system,
we observe that the symmetric part of $I_{\a\b}$ can be
removed by a local superfield Lorentz transformation acting on the
index $A$ in Eq.~(\ref{viellin}). In other words, its choice
corresponds to a residual local Lorentz symmetry of the linearized
SUGRA. To fix this symmetry completely we set
$I_{\a\b}+I_{\b\a}=0$. Then we obtain the unique solution,
\begin{gather}
\label{Malbe}
M_{\a\b}=\frac{1}{4}D_\a\bar
D^2L_\b+\frac{1}{8}\epsilon_{\a\b}D_\g\bar D^2L^\g\;,\\
\delta_L I_\a^{~\;\b}=-\frac{1}{8}\delta_\a^\b\,D_\g\bar D^2
L^\g\;,\quad
\delta_LI_{\a\dot\b}=0\;,\quad
\delta_LI_\a^{~\;b}=\frac{i}{2}\s^b_{\b\dot\b}D_\a(\bar
D^{\dot\b}L^\b-D^\b\bar L^{\dot\b})\;.
\label{Ialbe}
\end{gather}
From (\ref{Malbe}) using the structural relations (\ref{structrel})
we obtain the rotation matrix for Lorentz vectors,
\be
\label{Mab}
M_{ab}=\frac{1}{4}(\s_{ab})_\b^{~\;\a}D_\a\bar D^2L^\b
+\frac{1}{4}(\bar\s_{ab})^{\dot\a}_{~\;\dot\b}
\bar D^{\dot\b} D^2\bar L_{\dot\a}\;.
\ee
Note that its lowest component reads,
\be
\label{Mablowest}
M_{ab}\big|=(\s_{ab})^{\a\b}\lambda_{\a\b}-
(\bar\s_{ab})^{\dot\a\dot\b}\bar\lambda_{\dot\a\dot\b}\;,
\ee
which is precisely the matrix of local Lorentz transformations acting
on the physical tetrad, see Eq.~(\ref{comptrans2}).

Equations (\ref{Ialbe}) allow us to determine the vielbein components
entering them in terms of the fields $H_b$, $\Gamma$.
The unique combinations of these fields with the required transformation
properties are,
\be
\label{viel1}
I_\a^{~\;\b}=-\delta_\a^\b\,\frac{1}{2}\bar\Gamma'\;,~~~~~
I_{\a\dot\b}=0\;,~~~~~
I_\a^{~\;b}=-iD_\a H^b\;,
\ee
where
\be
\bar\Gamma'=-\frac{(3n\!+\!1)(n\!-\!1)}{4n}\,\bar\Gamma
-\frac{(3n\!+\!1)(n\!+\!1)}{4n}\,\Gamma
-\frac{(n\!+\!1)^2}{8n}\bar D^{\dot\a}D^\a H_{\a\dot\a}
+\frac{n^2\!-\!1}{8n}D^\a\bar D^{\dot\a}H_{\a\dot\a}\;.
\ee
Thus, the spinor covariant derivative of a scalar field takes the
form,
\be
\label{scalcovder}
{\cal D}_\a\Psi=\bigg(1+\frac{\bar\Gamma'}{2}\bigg)\,D_\a\Psi
+iD_\a H^b\,\d_b\Psi\;.
\ee
The expression for $\bar{\cal D}_{\dot\a}\Psi$ is obtained by complex
conjugation. Determination of the remaining components of the
linearized vielbein $I_a^{~\;\b}$, $I_a^{~\;b}$ requires invoking the
constraints on the superspace torsion imposed in non-minimal SUGRA
\cite{Girardi:1984eq}. This analysis is performed in
Appendix~\ref{app:torsion}.

For a superfield with spinor or vector indices the covariant
derivatives should be supplemented with a connection term,
\be
{\cal D}_A\Psi^B=E_A^{~\;M}\,\d_M\Psi^B+(-1)^{|A||B|}\Psi^C\,\Phi_{A
  C}^{\phantom{A C}B}\;,
\label{spincovder2}
\ee
where\footnote{Following \cite{WB}, we assume that the fields carrying even (odd)
  number of spinor indices take commuting (anti-commuting) values.}
\[
|B|=\begin{cases}
0,&\text{for}~~ B=b\\
1,&\text{for}~~B=\b~\text{or}~\dot\b
\end{cases}
\]
The connection coefficients $\Phi_{AC}^{\phantom{AC}B}$ obey the
structural relations of $SL(2)$ analogous to Eqs.~(\ref{structrel}),
\begin{gather}
\Phi_{A\g\b}=\Phi_{A\b\g}\;,\quad
\Phi_{\a\dot\g\dot\b}=-(\Phi_{\dot\a\g\b})^*
\;,\quad \Phi_{\dot\a\dot\g\dot\b}=-(\Phi_{\a\g\b})^*
\;,\quad \Phi_{a\dot\g\dot\b}=-(\Phi_{a\g\b})^*\;,\notag\\
\Phi_{Acb}=\frac{1}{2}\bar\s_c^{\dot\g\g}\bar\s_b^{\dot\b\b}
(\epsilon_{\dot\g\dot\b}\Phi_{A\g\b}-\epsilon_{\g\b}\Phi_{A\dot\g\dot\b})\;.
\label{Phistruct}
\end{gather}
In the linearized SUGRA the connections are expressed in terms of the
fields $H_b$, $\Gamma$. Below we will need explicit formulas for the
components $\Phi_{\dot\a C}^{\phantom{\dot\a C}B}$. These are found
simultaneously with the vielbein from the torsion constraints, see
Appendix~\ref{app:torsion}. The result reads,
\be
\label{connspin}
\Phi_{\dot\alpha\beta\gamma} = -\frac{1}{8} \big(\bar D^2D_{\beta}
H_{\gamma\dot\a}
+ \bar D^2 D_{\gamma} H_{\beta\dot\a}\big)\;,\qquad
\Phi_{\dot\alpha\dot\beta\dot\gamma} =
\frac{1}{2}\big(\epsilon_{\dot\alpha\dot\beta} \bar
D_{\dot\gamma}\Gamma
+ \epsilon_{\dot\alpha\dot\gamma} \bar D_{\dot \beta}\Gamma\big)\;.
\ee
Note that these formulas do not depend on the parameter $n$.
The connection in the vector representation is obtained from the
structural relation (\ref{Phistruct}),
\begin{equation}
\label{connvect}
\Phi_{\dot \alpha bc} = -\frac{1}{4}(\sigma_{bc})_\a^{~\;\beta}
\bar D^2 D^\a H_{\b\dot\alpha}
-(\bar\sigma_{bc})^{\dot\beta}_{~\;\dot\alpha} \bar D_{\dot\beta} \Gamma\;.
\end{equation}
Note that under the linearized super-diffeos the connections transform
as
\be
\label{varPhi}
\delta_L\Phi_{\dot\a C}^{\phantom{\a C}B}=-\bar D_{\dot\a} M_C^{~\;B}\;.
\ee
This is consistent with the transformations of the corresponding
covariant derivatives (\ref{spincovder2}) in the appropriate
representations of $SL(2)$.

\section{Breaking Lorentz invariance}
\label{sec:3}

\subsection{Perturbations of super-aether}
\label{sec:superaether}

We now want to generalize the linearized SUGRA to the case when
Lorentz invariance is broken down to the $SO(3)$ subgroup of spatial
rotations by a vacuum expectation value (VEV)
of a timelike vector field. To this end, we
introduce \cite{Pujolas:2011sk} a chiral vector superfield $U^a$
obeying the constraint (\ref{Unorm}). As a consequence of this
constraint, the field develops a c-number VEV $w^a$ satisfying the
relations,
\[
\Re w_a \Re w^a-\Im w_a\Im w^a=-1~,~~~~~
\Re w_a \Im w^a=0\;.
\]
They imply that $\Re w^a$ is always timelike and
$\Im w^a$ is spacelike. Thus, unless $\Im
w^a=0$, the vacuum
breaks both Lorentz and rotational
symmetries. In this paper we are interested in quadratic theory around
a rotationally invariant vacuum, so we focus on the case of real
$w^a$. Then there is a preferred Lorentz frame where $w^a$ has the
form,
\be
\label{wtimel}
w^a=(1,0,0,0)\;.
\ee
It is important to stress that, despite the breaking of Lorentz
invariance, all SUSY generators are preserved as they
corresponds to translations in the
superspace that leave the aether VEV invariant
\cite{GrootNibbelink:2004za,Bolokhov:2005cj,Pujolas:2011sk}.

Next, we expand the super-aether field about its VEV as in
Eq.~(\ref{Vdef}). 
As already mentioned in the Introduction, this expansion is forced on
us by the fact that we work with the linearized SUGRA. Indeed,
the latter is invariant under the linearized version of the local
Lorentz and coordinate transformations contained in the
parameter $L_\alpha$, see Sec.~\ref{sec:fieldcontent}. Let us look at
the original Einstein-aether theory (\ref{Sae}) and ask what sector of
it is invariant under these linearized transformations. Clearly, this
sector corresponds to
the quadratic part of the action expanded in metric and aether
fluctuations around the background. In particular, it would be
inconsistent to expand only in the metric while keeping the aether
nonlinear: the latter transforms non-trivially under the local
spacetime symmetries and the invariance would be lost. Returning to the
SUSY case, we conclude that the requirement of invariance with respect
to the linearized local SUSY requires restricting the equations
of motion (action) to the linear (quadratic) order in the super-aether
perturbation~$V^c$.

The constraint (\ref{Unorm}) expanded to linear order translates into
\be
\label{Vnorm}
w_a V^a=0\;,
\ee
whereas the chirality condition reads,
\begin{equation}
\label{Vchir}
\bar D_{\dot \alpha} V^c=-w^b\Phi_{\dot \alpha b}{}^c\;.
\end{equation}
Here we expanded the covariant derivative (\ref{spincovder2}) to
linear order both in aether perturbations and SUGRA fields.
Note that the explicit form of the connection (\ref{connvect}) implies
that it is chiral and hence the aether perturbation $V^c$ is linear,
$\bar D^2 V^c=0$.
The aether
perturbations transform non-trivially under
super-diffeomorphisms,
\begin{equation}
\label{Vtrans}
\delta_L V^a=w^bM_b^{~a}\;,
\end{equation}
where $M_b^{~a}$ is the matrix of Lorentz rotations (\ref{Mab}).

We define the components of the aether supermultiplet as
follows,
\be
\label{Vcomp}
v^a\equiv v^{R,a}+iv^{I,a}=V^a\big|~,~~~~~
\eta^a_\a={\cal D}_\a U^a\big|=(D_\a V^a+w^c\Phi_{\a c}{}^{a})\big|~,~~~~~
G^a=-\frac{1}{4}D^2 V^a\big|\,.
\ee
Note that in the definition of the aetherino $\eta^a_\a$ we used the
covariant spinor derivative acting on the full aether, which we
expanded to the linear order in the second equality. Due to this
definition, the aetherino is invariant under the linearized
super-diffeos, as it follows from the transformation laws
(\ref{Vtrans}) and (\ref{varPhi}). The same is true also for the
auxiliary field $G^a$ due to the property $D^2 M_b^{~a}=0$ implied by
the expression (\ref{Mab}). Thus, we have,
\bseq
\label{compVtrans}
\be
\label{deltaLetaG}
\delta_L\eta^a_\a=\delta_L G^a=0\;.
\ee
On the other hand, the lowest component $v^a$ transforms
non-trivially. From Eqs.\,(\ref{Mablowest}) and (\ref{lamxi})
we find,
\be
\delta_L v^a=-\frac{1}{2}w_b\,\big(\d^b\xi^{R,a}-\d^a\xi^{R,b}\big)\;
\label{deltaLva}
\ee
\eseq
The transformations of the super-aether components under
global SUSY are derived in Appendix~\ref{app:globalSUSY}.

The
component fields will be used in Secs.~\ref{sec:bosonic}, \ref{sec:fermionic}. Now
our goal is to find the most general superspace action quadratic in
the superfields $V^a$, $H_a$, $\Gamma$ and invariant under the
transformations
(\ref{Htrans}), (\ref{Gammatrans}), (\ref{Vtrans}).\footnote{It is
  worth noting that this action cannot be 
found simply by a linear gauging of the super-aether action of
\cite{Pujolas:2011sk}. Such gauging does not capture the terms
quadratic in the SUGRA fields $H_b$ and $\Gamma$ which are present in
the resulting super-aether action and are important
for the analysis of its physical content, see below.}

\subsection{Possible terms in the Lagrangian}
\label{sec:superterms}

First we notice that the only possible term in the superpotential is the
term enforcing the constraint (\ref{Vnorm}) by means of a chiral
Lagrange multiplier $\Lambda$
(cf. \cite{Pujolas:2011sk}),
\be
\label{soperpotential}
{\cal L}_{constr}=\int d^2\theta\; \Lambda w_a V^a +\text{h.c.}\;.
\ee
The combination $w_aV^a$ is chiral due to the relation (\ref{Vchir})
and anti-symmetry of the connection coefficient $\Phi_{\dot\a bc}$ in
the last two indices. No other chiral combination can be constructed
from $V^a$ and the SUGRA fields without using the spinor derivatives
$\bar D_{\dot\a}$. On the other hand, the terms in the superpotential that
involve $\bar D_{\dot\a}$ can be equivalently
written as contributions to the K\"ahler potential. For example,
\[
\int d^2\theta\, \bar D_{\dot\a}\Gamma  \bar D^{\dot\a}\Gamma\simeq
-2\int d^2\theta d^2\bar\theta\, \Gamma^2\;,
\]
and similarly for other contributions.
Here the sign $\simeq$ means `equal up to a total
derivative'. Thus, it is enough to consider the K\"ahler potential
only.

By analogy with (\ref{Sae}), we search for the action in the form
\[
S=\frac{1}{\vk^2}\int d^4x\int d^2\theta d^2\bar\theta\;\tilde{\mathbb{L}}\;,
\]
where the gravitational coupling with the mass dimension
$[\vk^{-2}]=2$ has been factored out in front of the Lagrangian.
All other parameters in the Lagrangian are assumed to be
dimensionless. This implies that the superspace Lagrangian $\tilde{\mathbb{L}}$
must be constructed from terms with zero mass dimension.
The dimensions of the object at our disposal are
\be
\label{dimensions}
[H_a]=-1~,~~~~[w^a]=[V^a]=[\Gamma]=0~,~~~~[D_\a]=[\bar
D_{\dot\a}]=1/2~,~~~~
[\partial_a]=[\Delta_a]=1\;.
\ee
Once the aether VEV
$w^a$ is included as the spurion to compensate for the breaking of the Lorentz
symmetry, the Lagrangian becomes a scalar with respect to global
Lorentz transformations.
Besides, the Lagrangian must be real.

As our goal is to find the most general super-aether action, we
proceed as follows. We first classify all inequivalent terms in the
quadratic Lagrangian satisfying the above requirements. In the next
subsection we will look
for their linear combinations invariant under the non-linearly
realized super-diffeomorphisms.

{\bf Operators quadratic in $V^a$.} We have a single operator in this class,
\be
\label{quadrV}
V_a\bar V^a\;.
\ee
The combination $V_aV^a$ and its complex conjugate can be rewritten
purely in terms of the SUGRA fields $H_a$ and $\Gamma$ using the
relation (\ref{Vchir}). Indeed,
\[
\int d^2\theta d^2\bar\theta\; V_aV^a
\simeq-\frac{1}{4}\int d^2\theta \bar D^2(V_aV^a)
=-\frac{1}{2}\int d^2\theta\; w^bw^c\Phi_{\dot\a ba}\Phi^{\dot\a~a}_{~c}\;.
\]
As discussed
above, the expression on the r.h.s. can be cast in the form of a
contribution into the K\"ahler potential.

{\bf Operators linear in $V^a$.} In total, there are four independent
combinations,
\be
\label{linV}
w^a V^b\d_a H_b~,~~~~ w^aV^b\d_b H_a~,~~~~
w^a V^b\d^cH^d\epsilon_{abcd}~,~~~~ V^aD^2 H_a\;,
\ee
plus their complex conjugate. From all other
terms the aether
perturbation can be eliminated by performing integration by parts and
using (\ref{Vchir}). For
example,
\begin{align}
&V^a\bar D^2H_a\simeq (\bar D^2V^a) H_a=0\;,\notag\\
&w^aV^b\Delta_a
H_b=w^aV^b\bigg(\!\!-\!i\d_a\!+\!\frac{1}{2}\bar\s^{\dot\a\a}_a\bar D_{\dot\a}D_\a\bigg)H_b
\simeq-iw^aV^b\d_aH_b
\!+\!\frac{1}{2}\bar\s^{\dot\a\a}_aw^a w^c\Phi_{\dot\a c}^{~~~b}D_\a H_b\;,\notag
\end{align}
and so on.

{\bf Operators without $V^a$.} This is the most numerous group of
terms. On the other hand, their role is in a sense auxiliary: as
discussed below, they are needed to compensate the variation of the
quadratic aether term (\ref{quadrV}) and do not give rise to any
independent Lagrangians. We relegate their classification to
Appendix~\ref{app:classify}. We find a total of $27$ terms that are
not equivalent to each other upon integration over the superspace.

\subsection{The invariant action}
\label{sec:invact}
The action for linearized SUGRA with broken Lorentz invariance is
obtained as a linear combination of the independent operators listed
in the previous subsection that is invariant under the gauge
transformations (\ref{Htrans}), (\ref{Gammatrans}), (\ref{Vtrans}).
To find this combination, let us analyze the variations of individual
operators.
We
start with the term quadratic in the aether perturbations,
\be
\label{V2trans}
\begin{split}
\delta_L \big(V_a & \bar V^a\big) = w^bM_{ba}\bar V^a+w^bM_{ba}V^a\\
\simeq& -w^bw^c L^\b(\s_{ba})_\b^{~\;\g}
 \bigg(\frac{1}{4}\bar D^2 \Phi_{\g c}^{\phantom{\g c}a}
+ i \d_{\g\dot\g}\Phi^{\dot\g~a}_{~c}\bigg)+\text{h.c.}\\
=&L^\beta \bigg(-\frac{i}{2}w^aw^b (\s_{ak})_\b^{~\;\g}\bar D^2D_\g\d^k H_b
 + \frac{i}{4}w^aw^b\bar D^2 D_\b\d_a H_b
-\frac{i}{8}\bar D^2 D_\beta \partial_k H^k\\
&~~~~~~
-iw^aw^b\sigma_{a\beta\dot\beta} \bar D^{\dot\beta} \partial_b \Gamma
-\frac{i}{4}\sigma_{k\beta\dot\beta} \bar D^{\dot\beta} \partial^k
\Gamma
- \frac{3}{16}\bar D^2 D_\beta \bar\Gamma\bigg)+\text{h.c.}\;,
\end{split}
\ee
where in the second line we used the chirality condition
(\ref{Vchir}). One observes that this variation is independent of
$V^a$ and is expressed exclusively in terms of the SUGRA fields $H_b$
and $\Gamma$. The final expression is at most quadratic in the spurion
field $w^a$. It also contains terms without the spurion
altogether. The same type of terms appear in the variation of the
Lorentz invariant supergravity operators. They come from the
contractions of the form $w_a w^a=-1$ which arise upon substitution of
the superspace connection into the second line.

On the other hand, the variations of the operators (\ref{linV})
contain contributions linear in $V^a$,
\bseq
\label{linVtrans}
\begin{align}
&\delta_L (w^aV^b\d_a H_b)\ni -\frac{1}{2}\bar
L_{\dot\b}\bar\s_b^{\dot\b\b}w^aD_\b\d_a V^b\;,\\
&\delta_L (w^aV^b\d_b H_a)\ni -\frac{1}{2}\bar
L_{\dot\b}\bar\s_a^{\dot\b\b}w^aD_\b\d_b V^b\;,\\
&\delta_L (w^aV^b\d^c H^d\epsilon_{abcd})\ni -\frac{1}{2}
\bar L_{\dot\b}\bar\s^{d\,\dot\b\b}w^aD_\b\d^c V^b\epsilon_{dacb}\;,\\
&\delta_L (V^a D^2H_a)\ni L_\b\big[-4i(\s_{ka})_\g^{~\;\b}D^\g\d^kV^a
+2iD^\b\d_aV^a\big]\;.
\end{align}
\eseq
It is straightforward to see that these cannot be canceled among
themselves or against 
variations of any other operators in the Lagrangian. Thus, we conclude
that the operators (\ref{linV}) do not appear in the invariant
action. 

The fact that there is only a single operator (\ref{quadrV}) with the
aether fluctuation $V^a$ which can enter the super-aether action
suggests that this action must be unique. Indeed, it would be
surprising to obtain independent Lorentz-violating Lagrangians that at
the quadratic level would consist purely of the gravitational
fields. We presently verify this intuition by explicitly solving the
constraints imposed by the linearized super-diffeos.  

To this aim we need to derive the variations of the $27$ ``auxiliary''
operators from Appendix~\ref{app:classify}. This technical task is
performed in Appendix~\ref{app:trans}. Let us note here that it is
simplified by classifying these operators according to their
transformation properties under the $R$-symmetry and $CP$. The
operators with different $R$-charges and $CP$-numbers do not mix under
the action of linearized super-diffeomorphisms and thus can be
considered separately.

For the $CP$-even sector the conditions that the variation of the
action under the super-diffeos cancels lead to a system of $17$ linear
equations for $17$ unknowns --- coefficients in front of the operator
(\ref{quadrV}) and $16$ operators (\ref{CPeven}). This system is
degenerate and has a general solution with two free parameters. One of
them is just the usual gravitational coupling $\vk^2$ and one recovers
the SUGRA action (\ref{NMact}) as part of the general solution. The
second free parameter can be chosen as the coefficient in front of the
operator (\ref{quadrV}) and the corresponding contribution into the
action reads,
\be
\label{superact}
\begin{split}
S_{\text{\AE}} =\frac{C}{2\vk^2}\int& d^4x d^4\theta\bigg[
 V_a \bar V^a + iw^aw^b\d_aH_b(\Gamma-\bar\Gamma)
+w^aw^b\DD_aH_b(\Gamma+\bar\Gamma)\\
&+\frac{1}{4}\big(\DD_kH_m\DD^kH^m-\d_kH_m\d^kH^m
-(\DD_mH^m)^2+(\d_mH^m)^2\big)\\
&+\frac{i}{4}\d_mH^m(\Gamma-\bar\Gamma)
+\frac{1}{4}\DD_mH^m(\Gamma+\bar\Gamma)
+\frac{3}{8}(\Gamma^2+\bar\Gamma^2)\bigg]\;,
\end{split}
\ee
where $C$ is a dimensionless coupling. This action is rather simple
and its invariance under
the linearized super-diffeos can be verified in a 
straightforward manner.
Curiously, it does not depend on the choice of the parameter $n$ labeling
the off-shell realizations of the non-minimal SUGRA. Note that, along
with explicitly Lorentz-violating terms in the first line, it contains
contributions that are Lorentz-invariant. There is no contradiction
here: these terms are needed to cancel the variations of the terms
from the first line which, as pointed out above, include
Lorentz-invariant contributions.  

Finally, the requirement of vanishing gauge variation in the $CP$-odd
sector leads to a system of 13 equations for only 8 unknowns which has
only a trivial solution (see
Appendix~\ref{app:trans}).
We conclude that (\ref{superact}) combined with (\ref{NMact}) gives
the most general action for supersymmetric aether coupled to
non-minimal linearized SUGRA
with linear compensator.

\section{Bosonic Lagrangian}
\label{sec:bosonic}
To understand the physical consequences of the action
(\ref{superact}), we compute the corresponding Lagrangian in
components. We start with the bosonic
part.

The components of the SUGRA fields $H_m$, $\Gamma$ and of the
super-aether $V^a$
have been introduced in
(\ref{Hcomp}), (\ref{compGamma}) and (\ref{Vcomp}), respectively.
We impose the Wess--Zumino gauge (\ref{WZ}) and the symmetry of the
tetrad (\ref{sime}).
Further, due to the
relation (\ref{weylfix}) fixing the Weyl invariance, the lowest
component of $\Gamma$ is not independent and is expressed through the
trace of the metric perturbation.
Applying the formula (\ref{LsupL}) to the superspace action (\ref{superact})
we obtain after a somewhat tedious, but straightforward calculation,
\be
\label{Laeth}
\begin{split}
{\cal L}^{\rm bos}_{\text{\AE}}=&\frac{C}{2\vk^2}
\bigg\{-\d_m v^a \d_m \bar v^a+G^a \bar G^a\\
&+v^{R,a} w^b\bigg(-2\epsilon_{abkm}\d_kd_m-\frac{1}{2}\d_a\d_kh_{kb}
+\frac{1}{2}\d_b\d_kh_{ka}+\d_bq^I_a-\d_a q^I_b
+\epsilon_{abkm}\d_k q_m^R\bigg)\\
&+v^{I,a}w^b\bigg(2\d_ad_b-2\d_bd_a
-\frac{1}{2}\epsilon_{abkm}\d_k\d_lh_{lm}
-\d_bq^R_a+\d_aq^R_b+\epsilon_{abkm}\d_kq_m^I\bigg)\\
&-\frac{1}{8}\mathbb{D}_{ca}\bar{\mathbb{D}}_{ca}
+\frac{n\!+\!1}{4(3n\!+\!1)}w^aw^b\big(h_{ab}\Box h-h\d_a\d_kh_{kb}\big)
+\frac{1}{2}w^aw^bq_m^I\big(\d_mh_{ab}-\d_ah_{mb}\big)\\
&+2w^aw^bd_b q_a^R+\frac{1}{2}w^aw^b h_{nb}\epsilon_{nakm}\d_k q_m^R
-\frac{3}{2}d_md_m-\frac{1}{8}h_{km}\Box h_{km}
-\frac{5}{32}\d_kh_{km}\d_lh_{lm}\\
&-\frac{2n+1}{8(3n+1)}h\d_k\d_lh_{kl}
+\frac{n(3n+2)}{8(3n+1)^2}h\Box h
-\frac{1}{8}q_m^I\d_kh_{km}-\frac{1}{4(3n+1)}q^I_m\d_m h\\
&+\frac{3}{8}q_m^Iq_m^I
-\frac{3}{8}q_m^Rq_m^R+\frac{1}{2}d_kq^R_k\bigg\}\;,
\end{split}
\ee
where
\be
\label{Dbb}
\begin{split}
\mathbb{D}_{ca}=&-w_c\bigg[q_a+2d_a-\frac{i}{2}\d_kh_{ka}
+i\frac{n}{3n+1}\d_ah\bigg]\\
&+\upeta_{ac}w^b\bigg[q_b+2d_b-\frac{i}{2}\d_kh_{kb}
+i\frac{n}{3n+1}\d_bh\bigg]\\
&+w^b\epsilon_{bmac}\bigg[iq_m-2id_m+\frac{1}{2}\d_kh_{km}
-\frac{n}{3n+1}\d_mh\bigg]\\
&-iw^b\d_bh_{ac}+iw^b\d_ah_{cb}
+w^b\epsilon_{bcmn}\d_mh_{na}-w^b\epsilon_{mnac}
\d_mh_{nb}\;.
\end{split}
\ee
Note that the parameter $C$ multiplies the kinetic term for the aether
perturbations in (\ref{Laeth}) and hence must be positive to ensure
the positivity of the kinetic energy.
The full bosonic action of the theory is obtained by adding
the standard supergravity Lagrangian (\ref{LSG}).

The next step is to integrate out the auxiliary fields. Clearly, the
fields $B$, $G^a$ simply vanish on the equations of motion.
On the other hand, the fields $d_m$, $q_m$ take non-zero values.
The result of integrating them out in the general case is rather
cumbersome and not illuminating. For the sake of clarity,
we will perform an explicit calculation under the assumption $C\ll
1$. To properly capture the mixing between the aether and gravity in the
first non-trivial order in $C$, we canonically
normalize the aether perturbations
so that their leading kinetic term becomes of order 1,
\[
v_a^{R,I}\mapsto \hat v_a^{R,I}=\sqrt{C}\, v_a^{R,I}\;.
\]
We are interested in the contributions to the Lagrangian through
order $O(C)$ in terms of the new fields.
To this end, it is sufficient to find the auxiliary
fields through order $O(\sqrt{C})$. However, later we will need also order
$O(C)$ contributions into the fields $d_m$ and $q_m^R$, so in deriving
them we go one order further. We obtain,
\bseq
\label{auxil}
\begin{align}
&d_m=\sqrt{C}\bigg[\frac{1}{2}w_m\d_a\hat v^{I,a}
-\frac{1}{2}w^a\d_a\hat v^I_m-\frac{1}{4}w^b\epsilon_{bkam}\d^k\hat
v^{R,a}\bigg]\notag\\
&~~~~~~~-\frac{C}{8}w^a w^d \epsilon_{abcm}\d^bh^{c}_{~d}+O(C^{3/2})\,,\\
&q^R_m=\sqrt{C}\bigg[\frac{n}{3n+1}w_m\d_a\hat v^{I,a}
-\frac{n}{3n+1}w^a\d_a\hat v^I_m-\frac{n+1}{2(3n+1)}w^b\epsilon_{bkam}\d^k\hat
v^{R,a}\bigg]\notag\\
&~~~~~~~-\frac{C(n+1)}{4(3n+1)}w^a w^d \epsilon_{abcm}\d^bh^{c}_{~d}
+O(C^{3/2})\;,\\
&q^I_m=\frac{1}{2}\d^kh_{km}-\frac{n}{3n\!+\!1}\d_m h
+\frac{\sqrt{C}n}{3n\!+\!1}\Big[w_m\d_a\hat v^{R,a}-w^b\d_b\hat v_m^R
-w^b\epsilon_{bkam}\d^k\hat v^{I,a}\Big]+O(C)\,.
\end{align}
\eseq
Substituting this back into (\ref{LSG}), (\ref{Laeth}) we arrive
at,
\be
\label{Ltot}
\begin{split}
{\cal L}_{SG}^{\rm bos}\!+\!{\cal L}_{\text{\AE}}^{\rm bos}=
\frac{1}{2\vk^2}&\bigg\{\frac{1}{4}h_{km}\Box h^{km}
+\frac{1}{2}\d^kh_{km}\d_lh^{lm}
-\frac{1}{2}\d_kh^{km}\d_mh+\frac{1}{4}\d_mh\d^mh\\
&-\d_m\hat v^{R}_a\d^m\hat v^{R,a}-\d_m\hat v^{I}_{a}\d^m\hat v^{I,a}
+\sqrt{C}\,\hat v^{R,a}w^b\big(\d_b\d^kh_{ka}-\d_a\d^kh_{kb}\big)\\
&-\frac{C}{4}w^aw^b\big(\d_ah_{mn}-\d_mh_{na}\big)
\big(\d_bh^{mn}-\d^mh^{n}_{~b}\big)
-\frac{C}{2} w^aw^b\d_a\hat v^{I,m}\d_b\hat v^I_m\\
&+\frac{C}{2}(\d_a\hat v^{I,a})^2
-Cw^bw^c\epsilon_{bkam} \d^k\hat v^{R,a}\d_c\hat v^{I,m}
+O(C^{3/2})\bigg\}\;.
\end{split}
\ee
where have made
a further rescaling of the fields,
\be
\label{vrescale}
\bigg(1+\frac{C\,n}{4(3n+1)}\bigg)\, \hat v^{R}_a\mapsto \hat
v^{R}_a~,~~~~~
\bigg(1-\frac{C\,n}{4(3n+1)}\bigg)\, \hat v^{I}_a\mapsto \hat
v^{I}_a\;.
\ee
One notices that the parameter $n$ has dropped from the
Lagrangian and,
apart from the usual gravitational coupling, the theory is described
by a single dimensionless constant $C$. When restricted to the case of
real aether, $\hat v^I_a=0$, the Lagrangian (\ref{Ltot}) coincides with the quadratic
part of the Einstein-aether Lagrangian (\ref{Sae}) for the choice of
couplings\footnote{In this comparison one should recall that $v^a$
  stands for the perturbation of the aether field in the {\em tetrad
    basis}. It is related to the perturbation in the tangent space
  by $\delta u^m=v^m-\frac{1}{2}h^{ma}w_a$.}
\be
\label{SGc}
c_1=C~,~~~~c_2=c_3=c_4=0\;.
\ee
Thus we conclude that the supersymmetrization of the Einstein-aether
model based on the embedding of aether into a chiral vector
supermultiplet reduces the number of free
parameters in the theory
from four down to one.

\section{Fermionic Lagrangian}
\label{sec:fermionic}

In principle, the fermionic part of the Lagrangian can also be
obtained from the superfield action (\ref{NMact}), (\ref{superact}) by the
application of Eq.~(\ref{LsupL}) and subsequent integration out of the
auxiliary fields. However, we will take a different route and adopt the
``on-shell'' formalism where one constructs the Lagrangian
directly in terms of
the dynamical fields: the metric perturbation $h_{mn}$, gravitino $\psi^m_\a$,
aether perturbation $v^a$ and aetherino $\eta^a_\a$. Apart from
leading more directly to the final answer, this approach will provide
us with a transparent
proof of uniqueness of the aether--supergravity coupling, not relying
on the rather tedious classification of the superfield operators in
Secs.~\ref{sec:superterms}, \ref{sec:invact}.

\subsection{Gauging the super-aether action}
\label{sec:gauging}

We start with the action of super-aether in flat spacetime
\cite{Pujolas:2011sk}, which at the quadratic level can be written as
\be
\label{SAEflat}
S_{\AE}^{\rm flat}=\frac{1}{2\vk^2}\int d^4x
\bigg[-\d_m\hat v^R_a\d^m\hat v^{R,a}-\d_m\hat v^I_a\d^m\hat v^{I,a}
-\frac{i}{2}\bar{\hat\eta}^a \bar\s^m\d_m\hat\eta_a\bigg]\;,
\ee
where we have rescaled the aetherino in the same way as the aether
perturbations,
$\hat\eta^a_\a=\sqrt{C}\, \eta^a_\a$. This action is invariant under
global spacetime translations and
global SUSY transformations,
\be
\label{flatSUSY}
\delta_G\hat
v_a^R=\frac{1}{2}(\zeta\hat\eta_a+\bar\zeta\bar{\hat\eta}_a)\;,
\quad
\delta_G\hat
v_a^I=-\frac{i}{2}(\zeta\hat\eta_a-\bar\zeta\bar{\hat\eta}_a)\;,
\quad
\delta_G\hat\eta_{a\a}=2i(\s^m\bar\zeta)_\a \d_m(\hat v^R_a+i\hat v^I_a)\,,
\ee
where $\zeta^\a$ is a coordinate independent parameter. These
transformations form a closed algebra on-shell, i.e. when the fields satisfy
the equations of motion,
\be
\label{flateom}
\Box\hat v^R_a=\Box\hat v^I_a=\bar\s^m\d_m\hat\eta_a=0\;.
\ee

Coupling to supergravity can be viewed as gauging of the above
symmetries \cite{VanNieuwenhuizen:1981ae}.
At linear order in the SUGRA fields, the coupling must
have the form,
\be
\label{lingauge}
{\cal L}_{\rm int}=\frac{1}{2}{\cal
  T}^{mn}h_{mn}
-{\cal S}^{m\a}\psi_{m\a}-\bar{\cal S}^m_{\dot\a}\bar\psi_m^{\dot\a}\;,
\ee
where ${\cal T}^{mn}$ is the symmetric super-aether energy-momentum tensor (EMT)
and ${\cal S}^{m\a}$ is the supercurrent corresponding to the
invariance under (\ref{flatSUSY}) via the Noether theorem. Both the EMT
and the supercurrent are conserved on-shell,
\be
\label{EMTScons}
\d_m{\cal T}^{mn}=\d_m{\cal S}^{m\a}=0\;.
\ee
Further, to allow for a consistent gauging, their SUSY transformations
have to be related on-shell as \cite{Komargodski:2010rb},
\bseq
\label{SCmult}
\begin{align}
\label{deltaGEMT}
&\delta_G{\cal T}_{mn}=-\frac{1}{2}\big(\zeta\s_{mk}\d^k {\cal S}_n
+\zeta\s_{nk}\d^k {\cal S}_m
+\bar\zeta\bar\s_{mk}\d^k \bar{\cal S}_n
+\bar\zeta\bar\s_{nk}\d^k \bar{\cal S}_m\big)\;,\\
\label{deltaGSma}
&\delta_G{\cal S}_{m\a}=-2i(\s^n\bar\zeta)_\a{\cal T}_{nm}
-4\epsilon_{mkln} (\s^k\d^l\bar\Xi^n)_\a\;.
\end{align}
\eseq
Here $\Xi^n_\a$ is the spin-vector appearing as the leading correction to the
on-shell SUSY transformation of gravitino.
For completeness we review the
derivation of Eqs.~(\ref{SCmult}) in Appendix~\ref{app:supercurrent}.

The coupling (\ref{lingauge}) contributes into the quadratic action if
the EMT and supercurrent contain terms
linear in $\hat v_a^{R,I}$, $\hat\eta_{a\a}$. On the other
hand, the Noether procedure applied to the action (\ref{SAEflat})
yields expressions quadratic in these fields. Thus, the linear terms
in the EMT and supercurrent must be of pure ``improvement'' type
(see e.g. \cite{Weinberg}),
\be
\label{improve}
{\cal T}^{mn}=\d_k{\cal M}^{km\,n}~,~~~~~~~
{\cal S}^{m\a}=\d_k{\cal N}^{km\,\a}\;,
\ee
where the differentiated (spin-)tensors are anti-symmetric in the first
pair of indices,
\be
\label{MNantisym}
{\cal M}^{km\,n}=-{\cal M}^{mk\,n}~,~~~~~~~
{\cal N}^{km\,\a}=-{\cal N}^{mk\,\a}\;.
\ee
It is worth stressing that (\ref{improve}) are on-shell equations and may or
maynot hold off-shell.
Our task now is to work out the most general form of the linearized
aether EMT and the supercurrent.

Let us start with the on-shell
EMT. The tensor ${\cal M}^{km\,n}$ must be constructed from terms
that are linear in the aether perturbations $\hat v_a^R$ or $\hat
v_a^I$, can contain one or several insertions of the VEV $w^c$
and, on dimensional grounds, must include a single derivative
$\d_n$.  Recalling the orthogonality of $w^c$ and
$\hat v_a$, we arrive to the following linear combination,
\be
\label{Mgen}
\begin{split}
{\cal M}^{km\,n}=&A_1(w^k\d^m\hat v^{R,n}-w^m\d^k\hat v^{R,n})
+A_2(w^k\d^n\hat v^{R,m}-w^m\d^n\hat v^{R,k})\\
&\!+A_3(w^n\d^k\hat v^{R,m}-w^n\d^m\hat v^{R,k})
+A_4(\upeta^{kn}w^m\d_a\hat v^{R,a}-\upeta^{mn}w^k\d_a\hat v^{R,a})\\
&\!+A_5(\upeta^{kn}w^a\d_a\hat v^{R,m}-\upeta^{mn}w^a\d_a\hat v^{R,k})
+A_6(w^k w^n w^a\d_a\hat v^{R,m}-w^m w^n w^a\d_a\hat v^{R,k})\\
&\!+A_7\,\epsilon^{kmna}w_a\d_b\hat v^{R,b}
+A_8\,\epsilon^{kmna}w^b\d_b\hat v^{R}_a
+A_9\,\epsilon^{kmab}w_a\d^n\hat v_b^R\\
&\!+A_{10}(\epsilon^{knab}w^m \d_a\hat v_b^R-\epsilon^{mnab}w^k \d_a\hat
v_b^R)
+A_{11}(\epsilon^{knab}w_a \d^m\hat v_b^R-\epsilon^{mnab}w_a \d^k\hat
v_b^R)\\
&\!+A_{12}(\epsilon^{knab}w_a \d_b\hat v^{R,m}
-\epsilon^{mnab}w_a \d_b\hat v^{R,k})
+A_{13}(\upeta^{kn}\epsilon^{mabc}
-\upeta^{mn}\epsilon^{kabc})w_a \d_b\hat v^R_c\\
&\!+A_{14}\,\epsilon^{kmab}w^n w_a w^c\d_c\hat v_b^R
+ \text{terms with}~ \hat v^I_a \;,
\end{split}
\ee
where $A_i$, $i=1,\ldots,14$, are dimensionless coefficients.
We show explicitly only the terms containing $\hat v^R_a$,
the part with $\hat v^I_a$ has similar form with 14 more parameters $\tilde
A_i$.
In
deriving this expression we omitted the terms that vanish identically
upon taking the divergence and simplified the part
with three vectors $w^c$ by using
the identity,\footnote{To prove it one notices that the combination on the l.h.s.
is totally antisymmetric in the indices $k\,l\,m\,n$. On the other hand, its
contraction with the vector $w^k$ vanishes. In four dimensions
a tensor with such properties is identically zero.}
\be
\label{epsident}
w^k w_a\epsilon^{almn}-w^l w_a\epsilon^{amnk}+w^m w_a\epsilon^{ankl}
-w^n w_a\epsilon^{aklm}+\epsilon^{klmn}=0\;.
\ee
Eq.~(\ref{Mgen}) is quite lengthy. However, the
number of independent terms is drastically reduced by imposing that ${\cal T}^{mn}$
obtained from ${\cal M}^{km\,n}$ must be symmetric,\footnote{Note that
${\cal M}^{km\,n}$ itself need not, and actually cannot, be symmetric
in the indices $m$ and $n$.
}
${\cal T}^{mn}={\cal T}^{nm}$.
This leaves only six free parameters: $A_1$, $A_3$, $A_{11}$ and their
tilded counterparts. The resulting EMT reads,
\be
\label{Tmnonshell}
\begin{split}
{\cal T}^{mn}_\text{on-shell}&=A_1 (w^a\d_a\d^m\hat v^{R,n}+w^a\d_a\d^n\hat v^{R,m})
-A_3(w^n\d^m \d_a\hat v^{R,a}+w^m\d^n \d_a\hat v^{R,a})\\
&-(A_1-A_3)\upeta^{mn} w^a\d_a\d_b\hat v^{R,b}
+A_{11}(\epsilon^{mabc}w_a\d^n\d_b\hat
v_c^R+\epsilon^{nabc}w_a\d^m\d_b\hat v_c^R)\\
&+\text{terms with}~ \hat v^I_a
\end{split}
\ee
where we have omitted terms proportional to the equations of motion
(\ref{flateom}).

So far, the analysis was restricted on-shell. The off-shell expression
for ${\cal T}^{mn}$ can contain in addition to (\ref{Tmnonshell}) two more terms,
\be
\label{DeltaTmn}
\Delta {\cal T}^{mn}=-A\, (w^m\Box\hat v^{R,n}+w^n\Box\hat v^{R,m})
-\tilde A\, (w^m\Box\hat v^{I,n}+w^n\Box\hat v^{I,m})\;.
\ee
On the other hand, one obtains further constraints by requiring the
off-shell invariance of the aether Lagrangian under linearized
diffeomorphisms. With respect to the latter the metric perturbation $h_{mn}$
transforms in the standard way (\ref{standlocal}), whereas for the
aether perturbations we have (cf. Eq.~(\ref{deltaLva})),
\be
\label{hatvdiff}
\delta_L\hat v^R_a=-\frac{\sqrt{C}}{2}w^b(\d_b\xi^R_a-\d_a\xi^R_b)~,
~~~~~~~\delta_L \hat v^I_a=0\;.
\ee
Assuming, as before, that $C$ is a small parameter, the sum of the
flat-space action (\ref{SAEflat}) and the interaction term
(\ref{lingauge}) must be invariant at order $O(\sqrt{C})$. Assuming
further that all coefficients in the EMT are of order $\sqrt{C}/(2\vk^2)$, we
get the relations,
\[
A_1=\frac{\sqrt{C}}{2\vk^2}+A~,~~~~
A_3=\frac{\sqrt{C}}{2\vk^2}-A~,~~~~\tilde A_1=-\tilde A_3=\tilde
A~,~~~~ A_{11}=\tilde A_{11}=0\;.
\]
Thus, the general off-shell expression for the linearized EMT reads,
\be
\label{Tmnoffshell}
\begin{split}
{\cal T}^{mn}&=\frac{\sqrt{C}}{2\vk^2} \big(w^c\d_c\d^m\hat v^{R,n}
-w^n\d^m \d_c\hat v^{R,c}\big)\\
&+\!A\,\big(w^c \d_c \d^m\hat v^{R,n}
\!+\!w^n\d^m \d_c\hat v^{R,c}
\!-\!\upeta^{mn} w^c\d_c\d_b\hat v^{R,b}
\!-\!w^m\Box\hat v^{R,n}\big)\\
&+\!\tilde A\,\big(w^c \d_c \d^m\hat v^{I,n}
\!+\!w^n\d^m \d_c\hat v^{I,c}
\!-\!\upeta^{mn} w^c\d_c\d_b\hat v^{I,b}
\!-\!w^m\Box\hat v^{I,n}\big)+(m\leftrightarrow n)\,.
\end{split}
\ee
We observe that the terms proportional to $\sqrt{C}$ yield, upon
integration by parts,
precisely the aether--metric mixing in the Lagrangian
(\ref{Ltot}).
One can check that the extra parameters $A$ and $\tilde A$
correspond to non-minimal couplings of the complex
aether to the Ricci tensor $R_{mn}$ of the form,
\[
R_{mn}u^m \bar u^n~,~~~~ R_{mn}(u^m u^n+\bar u^m\bar u^n),~~~~
i R_{mn}(u^m u^n-\bar u^m\bar u^n)\;,
\]
where $u^m=w^m+v^{R,m}+iv^{I,m}$ is the full non-linear aether
field. Only two free parameters appear at the linearized level because
the first two operators have the same expansion at the quadratic
order.

Up to now, we have not imposed any restrictions due to
supersymmetry. To do this, we construct the general linear supercurrent of
the form (\ref{improve}) and require that it is related on-shell to
the transformation of the EMT by Eq.~(\ref{deltaGEMT}). This will turn
out to be sufficient to completely fix the form of the current and
EMT: in particular, we will not need to use the second
Eq.~(\ref{deltaGSma}). The spin-tensor ${\cal N}^{km\,\b}$ must be
linear in the aetherino field and, by dimensionality, cannot contain
derivatives. Guided by the form of the EMT (\ref{Tmnoffshell}), we
consider only terms linear in $w^a$. Recall also
that $w^a$ and $\hat \eta^a_\b$ are orthogonal. We are left with four
possible operators,\footnote{In this derivation we eliminated the term
$\epsilon^{k}_{~abc}w^a\s^{bm}\hat\eta^{c\b}-(k\leftrightarrow m)
$
by using the identity
$\s^{bm}=(i/2)\epsilon^{bmnl}\s_{nl}$.
}
\be
\label{Ngen}
\begin{split}
{\cal N}^{km\,\b}=&a_1\,(w^k\hat\eta^{m\b}-w^m\hat\eta^{k\b})
+a_2\,(w^k\s^{ma}\hat\eta_{a}^{\b}-w^m\s^{ka}\hat\eta_{a}^{\b})\\
&+a_3\,(w_a\s^{ak}\hat\eta^{m\b}-w_a\s^{am}\hat\eta^{k\b})
+a_4\epsilon^{kmab}w_a\hat\eta_{b}^{\b}\;,
\end{split}
\ee
with free coefficients $a_{1,\ldots,4}$. The resulting on-shell
current reads,
\be
\label{SConshell}
\begin{split}
{\cal
  S}^{m\b}_\text{on-shell}=&\Big(a_1+\frac{a_3}{2}\Big)w^k\d_k\hat\eta^{m\b}
-\Big(a_1-\frac{a_2}{2}\Big)w^m\d_k\hat\eta^{k\b}\\
&+a_2w^k\s^{ma}\d_k\hat\eta_a^\b-a_3w_a\s^{am}\d_k\hat\eta^{k\b}
+a_4\epsilon^{mabc}w_a\d_b\hat\eta_c^\b\;.
\end{split}
\ee
We substitute it into the relation (\ref{deltaGEMT}), use equations of
motion to simplify the result and compare the coefficients in front of
independent terms. We find that Eq.~(\ref{deltaGEMT}) can be satisfied
only if
\[
A=\tilde A=a_2=a_3=a_4=0~,~~~~~a_1=-\frac{\sqrt{C}}{\vk^2}\;.
\]
We conclude that the linear supercurrent is unique and has the form,
\be
\label{SCoffshell}
{\cal S}^{m\b}=-\frac{\sqrt{C}}{\vk^2}
\big(w^c\d_c\hat\eta^{m\b}-w^m\d_c\hat\eta^{c\b}\big)\;.
\ee
This implies the uniqueness of the aether coupling to linear
supergarvity. Notice that the above supercurrent is conserved identically
(not only on-shell). Therefore, the gravitino coupling in
Eq.~(\ref{lingauge}) is automatically invariant under the local SUSY
transformations (\ref{standlocal}). We presently discuss its invariance with
respect to the global SUSY.

\subsection{On-shell supersymmetry}
\label{sec:onshell}

The interaction (\ref{lingauge}) gives rise to terms of order $O(\sqrt{C})$
in the total action. We saw that in the bosonic Lagrangian the next
corrections come at order $O(C)$. In principle, the same could happen
in the fermionic part. However, we now argue that this is not the
case and the total fermionic Lagrangian through order $O(C)$ reads,
\be
\label{Sferm}
\begin{split}
&{\cal L}_{SG}^{\rm ferm}+{\cal L}_{\AE}^{\rm ferm}=
\frac{1}{2\vk^2}\bigg\{8\epsilon^{klmn}\bar\psi_k\bar\s_l\d_m\psi_n
-\frac{i}{2}\bar{\hat\eta}^k\bar\s^m\d_m\hat\eta_k\\
&\qquad\,+2\sqrt{C}\,w^c(\psi^k\d_c\hat\eta_k-\psi_c\d_k\hat\eta^k)
+2\sqrt{C}\,w^c(\bar\psi^k\d_c\bar{\hat\eta}_k-\bar\psi_c\d_k\bar{\hat\eta}^k)
+O(C^{3/2})\bigg\}.
\end{split}
\ee
To show this we need to work out the global SUSY transformations of
the physical component fields. These are obtained by substituting
the on-shell
values of the auxiliary fields
into
the general formulas of Appendix~\ref{app:globalSUSY}.

We start with the gravity sector. The on- and off-shell
transformations of the metric coincide and are given by
Eq.~(\ref{tildedGh}). For gravitino we use Eq.~(\ref{tildedGpsi}),
where we substitute $B=0$ and $d_m$, $q_m^R$ from
Eqs.~(\ref{auxil}). This yields,
\be
\label{onshelldpsi}
\begin{split}
\tilde\delta_G \psi_{m\a}=&\frac{1}{2}(\s^{kn}\zeta)_\a \d_k h_{nm}\\
&+\sqrt{C}\bigg[-\frac{i}{4} w^c(\s_c\bar\s_m\zeta)_\a \d_b\hat v^{I,b}
+\frac{i}{4} w^c(\s_b\bar\s_m\zeta)_\a \d_c\hat v^{I,b}
-\frac{i}{4}\zeta_\a w^c\epsilon_{cklm}\d^k\hat v^{R,l}\bigg]\\
&-\frac{iC}{8}\zeta_\a w^c w^b \epsilon_{cklm}\d^k h^l_{~b}
+O(C^{3/2})\;.
\end{split}
\ee
One can check that the term of order $O(\sqrt{C})$ here
matches the extra piece $\Xi_{m\a}$ in the on-shell transformation
(\ref{deltaGSma})
of the
supercurrent (\ref{SCoffshell}).

The aether sector requires more work. First, from
Eq.~(\ref{tildedGv}) one observes that to get the transformation of $\hat v_a^{R}$
to order $O(C)$ one needs to know the fermionic auxiliary field
$\omega_\a$ through order $O(\sqrt{C})$. In Appendix~\ref{app:omega}
we describe how this field can be found using the superspace equations
of motion. The result reads,
\be
\label{omega}
\omega_\a=\frac{\sqrt{C}n}{3n+1} w^b (\s_{bc}\hat\eta^c)_\a+O(C)\;.
\ee
Second, one should remember the rescaling of the aether components at
order $O(C)$, Eq.~(\ref{vrescale}). In addition, it turns
convenient to redefine the aetherino field as follows,
\be
\label{etarescale}
\bigg(1-\frac{C}{4(3n+1)}\bigg)\,\hat\eta_{a\b}
+\frac{Cn}{2(3n+1)}P_a^b\, (\s_{bc}\hat\eta^c)_\b \mapsto \hat\eta^a_\b\;,
\ee
where
\be
\label{projector}
P^b_a\equiv w^bw_a+\delta^b_a
\ee
is the projector on the
hyperplane orthogonal to $w^a$.
Collecting everything together and substituting into
Eqs.~(\ref{tildedGV}) we arrive at
\bseq
\label{dVosh}
\begin{align}
\label{dvRosh}
\tilde\delta_G\hat v^R_a=&\frac{1}{2}\zeta\hat\eta_a
+\sqrt{C}w^c (i\zeta\s_c\bar\psi_a-i\zeta\s_a\bar\psi_c)
+\frac{Cn}{4(3n+1)}P^b_a\, \zeta\s_{bc}\hat\eta^c+{\rm h.c.}\;,\\
\label{dvIosh}
\tilde\delta_G\hat v^I_a=&-\frac{i}{2}\zeta\hat\eta_a
+\frac{iCn}{4(3n+1)}P^b_a\, \zeta\s_{bc}\hat\eta^c+{\rm h.c.}\;,\\
\tilde\delta_G\hat\eta_{a\b}=&2i(\s^m\bar\zeta)_\b\d_m\hat v_a
+i\sqrt{C}w^c (\s^k\bar\zeta)_\b [\d_c h_{ak}-\d_a h_{ck}]
-iCw^b w^c\epsilon_{cmna} (\s^m\bar\zeta)_\b\d_b\hat v^{I,n}\notag\\
&+\frac{Cn}{2(3n+1)}\big[-i P^c_a(\s^m\bar\zeta)_\b\d_c\bar{\hat v}_m
+iP^c_a(\s_c\bar\zeta)_\b\d_m\bar{\hat v}^m
+P^c_a\epsilon_{ckmn}(\s^k\bar\zeta)_\b\d^m\bar{\hat v}^n\big]\;,
\label{dvetaosh}
\end{align}
\eseq
where we have used the
identity (\ref{epsident}) to simplify the $\hat\eta_{a\b}$
variation. Corrections to these expressions are of order $O(C^{3/2})$.

It is now a matter of a straighforward calculation to verify that the
Lagrangian given by the sum of (\ref{Ltot}) and (\ref{Sferm}) is
invariant under SUSY transformations (\ref{tildedGh}), (\ref{onshelldpsi}),
(\ref{dVosh}) through order $O(C)$. We leave it as an exercise to
the reader.

We have also checked that the SUSY algebra closes on shell, up to the
gauge transformations (\ref{standlocal}), (\ref{deltaLva}). Namely, the
commutator of two SUSY transformations with parameters $\zeta_1$ and
$\zeta_2$ reads,
\bseq
\begin{align}
&[\tilde\delta_{G1},\tilde\delta_{G2}]h_{mn}=-2 Z^k\d_k h_{mn}+
\d_m\xi_n^R+\d_n\xi^R_m\;,\\
&[\tilde\delta_{G1},\tilde\delta_{G2}]\psi_{m\a}=-2Z^k\d_k\psi_{m\a}
+\d_m\varepsilon_{\a}+\text{e.o.m.}\;,\\
&[\tilde\delta_{G1},\tilde\delta_{G2}]\hat v_a^R=-2 Z^k\d_k\hat v_a^R
-\frac{\sqrt{C}}{2} w^b(\d_b\xi^R_a-\d_a\xi^R_b)\;,\\
&[\tilde\delta_{G1},\tilde\delta_{G2}]\hat v_a^I=-2 Z^k\d_k\hat
v_a^I\;,\\
&[\tilde\delta_{G1},\tilde\delta_{G2}]\hat\eta_{a\beta}=-2Z^k\d_k\hat\eta_{a\beta}
+\text{e.o.m.}\;,
\end{align}
\eseq
where
\be
Z^k=i(\zeta_1\sigma^k\bar\zeta_2-\zeta_2\s^k\bar\zeta_1)~,~~~~
\xi^R_n=Z^k h_{kn}\;,
\ee
and the fermionic gauge parameter
$\varepsilon_\a$ is linear in the fields $\psi_{m\a}$,
$\hat\eta_{a\beta}$; we do not write its explicit expression as it is
rather cumbersome. The terms denoted by
 ``e.o.m.'' vanish on
the fermionic equations of motion obtained from the
Lagrangian (\ref{Sferm}).
This proves that the fermionic action (\ref{Sferm}) is complete as any
additional terms of order $O(C)$
would spoil the supersymmetry.

Notice the following peculiarity. While the component Lagrangians
(\ref{Ltot}), (\ref{Sferm}) are independent of the non-minimal SUGRA
parameter $n$, the transformations (\ref{dVosh}) contain
$n$-dependent pieces. These pieces cancel among each other in the
variation of the total Lagrangian. This corresponds to an additional
supersymmetry of
the flat-space aether action (\ref{SAEflat}), besides the standard
SUSY (\ref{flatSUSY}). Coupling to SUGRA breaks the
extra SUSY and ties it to the first one. The preserved linear
combination of symmetry generators is
different for different $n$. We do not know if this leads
to the dependence of the Lagrangian on $n$ at higher orders in
$C$ or such dependence can always be eleiminated by a field
redefinition.
Investigating this issue goes beyond the scope of the present
paper.

\section{Physical implications}
\label{sec:physical}

\subsection{Particle spectrum, enhancement of graviton multiplet}
\label{sec:spectrum}

Let us discuss the spectrum of modes described by the Lagrangians
(\ref{Ltot}), (\ref{Sferm}). We will work in the frame where the VEV $w^a$ is purely
timelike as given by (\ref{wtimel}), so that the rotational symmetry
is preserved. As the Lagrangian is quadratic both in space- and
time-derivatives, all modes have linear dispersion relations,
\be
\label{lindisp}
E=s\cdot p\;,
\ee
where $E$ and $p$ are the energy and the absolute value of the mode's
momentum, $s$ is the mode's velocity.
Due to invariance with respect to spatial rotations, the
modes are also characterized, as in the familiar Lorentz invariant
case, by the projection of the angular momentum on the direction of
motion, i.e. helicity. The maximal helicity present in the spectrum is
$2$, which corresponds to the transverse-traceless excitations of the
metric --- gravitons. It is straightforward to see from (\ref{Ltot})
that the corresponding squared velocity is (cf. \cite{Jacobson:2004ts}),
\be
\label{sgrav}
s^2_{h=2}=\frac{1}{1-C}\approx 1+C\;,
\ee
which differs from $1$ whenever Lorentz invariance is broken ($C\neq
0$). This is in contrast with the situation for chiral and gauge
SUSY multiplets which, under broad assumptions, retain unit
propagation velocity even in the presence of Lorentz symmetry breaking
\cite{GrootNibbelink:2004za,Bolokhov:2005cj}.

The deviation of the graviton velocity from one has interesting
consequences for the structure of the gravitational
supermultiplet. Indeed, consider the representation of the SUSY
algebra corresponding to the dispersion relation of the form
(\ref{lindisp}). Following the standard procedure \cite{WB} one
rotates the direction of the particle momentum to align with the third
axis. With this choice, the anticommutators of the supercharges take
the form,
\be
\label{antic}
\{Q_\a,\bar Q_{\dot\b}\}=2E
\begin{pmatrix}
1+s^{-1}&0\\
0&1-s^{-1}
\end{pmatrix}
\ee
Note that unitarity requires the anticommutator of two conjugate
operators to be non-negative. Comparing with (\ref{antic}) we conclude
that in supersymmetric
theories the velocity of particles is always greater or equal to
one,
$s\geq 1$. If $s=1$, as it happens, in particular, in the standard
Lorentz invariant case, the lower right element in the above matrix is
zero implying that one pair of the supercharges vanish identically,
$Q_2=\bar Q_{\dot 2}=0$. The other pair of the supercharges $Q_1$,
$\bar Q_{\dot 1}$ describes fermionic annihilation and creation
operators. Thus, starting from the state with the lowest helicity $h$,
annihilated by $Q_1$, one can create a single state with helicity
$(h+1/2)$ by applying $\bar Q_{\dot 1}$.
As a consequence, in the Lorentz invariant case the
gravitational multiplet consists of just two states with helicities
$h=-2$ (graviton) and $h=-3/2$ (gravitino).\footnote{The states with
  opposite helicities $+2$, $+3/2$ appear upon the CPT conjugation.}
However, whenever $s>1$, the anticommutator of $Q_2$ and $\bar
Q_{\dot 2}$ does not vanish and they form an independent pair of
creation-annihilation operators. It implies that the multiplet must
contain two additional states: one with $h=-3/2$ and another with $h=-1$.
We conclude that the gravitational multiplet gets enhanced.

In the
model of this paper the additional states come from the aether
superfield. Indeed, its aetherino component $\hat\eta^m_\a$
carries both a spinor and a vector index and decomposes into a pair
of $h=\pm 3/2$ states and two pairs of $h=\pm 1/2$ states. The above
reasoning implies that the first pair is absorbed by the graviton
multiplet. The aether itself, represented by $\hat v^{R,I}_m$,
contains two pairs of $h=\pm 1$ states and a pair of $h=0$ states.
One of the $h=\pm 1$ pairs must join
the graviton multiplet. We verify this by computing the velocities of
the fermionic and bosonic modes in Appendix~\ref{app:disph1}. As
expected, we find
that for the helicity-3/2 modes and one pair of helicity-1 modes the
velocities coincide with that of gravitons,
\be
\label{svect1}
s^2_{h=3/2}=s_{h=1,(1)}^2=1+C\;.
\ee
We identify these modes as belonging to the graviton multiplet. On the
other hand, the remaining pair of helicity-1 modes, the helicity-1/2
modes and helicity-0 modes have unit velocities,\footnote{Up to
  possible corrections of order $O(C^2)$ that
  we neglect in our analysis.}
\be
\label{svect2}
s_{h=1,(2)}^2=s^2_{h=1/2}=s^2_{h=0}=1\;.
\ee
Thus, apart from the graviton multiplet, the theory
contains 4 bosonic degrees of freedom propagating with unit velocity
that match the two pairs of
$h=\pm 1/2$ fermionic states contained in $\hat\eta^m_\a$.

\subsection{Comments on phenomenology}
\label{sec:pheno}

As discussed in Refs.~\cite{GrootNibbelink:2004za,Bolokhov:2005cj,Pujolas:2011sk},
supersymmetry suppresses a direct coupling of aether to the Standard
Model fields. As a result, the non-gravitational dynamics of the
Standard Model sector is essentially relativistic. In particular, the
electromagnetic waves propagate with unit velocity. 

Thus, all constraints on the model come from the gravitational
sector. We observe
that at low energies SUSY must be broken which gives mass to the
imaginary part of the aether perturbations \cite{Pujolas:2011sk}. The
precise value of the mass depends on the SUSY breaking pattern;
nevertheless generically one expects the corresponding Compton wavelength
to be much shorter than astronomical scales. Then at
these scales the imaginary part of the aether is irrelevant and the
model
reduces to the (linearized) Einstein-aether
theory with the parameters (\ref{SGc}).\footnote{SUSY breaking can, in principle,
  introduce deviations from the values (\ref{SGc}) in the low-energy
  theory. However, these deviations are small if the SUSY
  breaking scale in the aether sector lies hierarchically below the scale of Lorentz
  symmetry breaking set by $M_*=\vk^{-1}\sqrt{C}$. We assume that this is
  the case.}
Hence we can use the constraints on the general
Einstein-aether theory \cite{Oost:2018tcv} by restricting them to the case
(\ref{SGc}). Let us briefly review them.

We have seen that the velocity of gravitons, and hence gravity waves,
in the super-aether model necessarily exceed
unity.
Detection of the gravity wave signal from the neutron star
merger GW170817 \cite{TheLIGOScientific:2017qsa}
in coincidence with the electromagnetic counterpart
\cite{Goldstein:2017mmi,Savchenko:2017ffs}
places a stringent bound on such deviation
\cite{Monitor:2017mdv}. This translates into the limit
on the model parameter~$C$,
\be
\label{CWlimit}
C<1.4\times 10^{-15}~~~~~~~~~~~~\text{(GW170817)}\;.
\ee
This is the strongest constraint on the model to date.\footnote{
A very strong indirect lower bound on the graviton velocity follows
from the absence of gravitational Cherenkov radiation by ultra-high
energy cosmic rays \cite{Moore:2001bv}. However, it is automatically
satisfied for
superluminal gravitons.}
It can be viewed as the limit on the energy scale of the
Lorentz symmetry violation, $M_*\equiv\vk^{-1}\sqrt{C}\lesssim 10^{11}$\,GeV.

Independent, though weaker, limits come from the tests
of general relativity within the Solar System and from the
observations of solitary pulsars. The Solar System tests place bounds
on the values of the post-Newtonian parameters $\a_1$, $\a_2$
describing deviations from Lorentz invariance~\cite{Will:2005va},
\be
\label{SScons}
|\a_1|\lesssim 10^{-4}~,~~~~|\a_2|\lesssim 10^{-7}\;.
\ee
The post-Newtonian parameters for the Einstein-aether model were derived in
\cite{Foster:2005dk}; for the choice (\ref{SGc}) they reduce to
\be
\label{alphasC}
\a_1=0~,~~~~~\a_2=-\frac{2C}{2-C}\;.
\ee
Hence the bound (\ref{SScons}) translates into
\be
\label{SSCcons}
C\lesssim 10^{-7}~~~~~~~~~~~~\text{(Solar System)}\;.
\ee
A more stringent bound
\be
\label{alphahatcons}
|\hat \a_2|<1.6\times 10^{-9}
\ee
on the analog of the parameter $\a_2$ for strong gravitational field
has been obtained in \cite{Shao:2013wga} by analyzing the dynamics of
solitary pulsars. Strictly speaking, application of this bound to our
model requires its non-linear generalization which is beyond the scope
of the present work. However, due to the uniqueness of the
Einstein-aether theory, this non-linear generalization must reduce to
it below the SUSY breaking scale, with the SUSY origin of the theory
still being encoded in the values (\ref{SGc}) of the parameters. The
relations between the strong- and weak-field parameters in the
Einstein-aether theory have been derived in the
Ref.~\cite{Yagi:2013ava}.
In general, they involve the sensitivities
characterizing the change in the binding energies of neutron stars
due to their motion with respect to the preferred frame. These
depend on the masses of the stars which complicates the translation of
the bound (\ref{alphahatcons}) into constraints on the model parameters.
However, for the choice
(\ref{SGc}) the sensitivities drop out of the relation between $\hat
\a_2$ and $\a_2$ and one gets simply
$\hat\a_2=\a_2$. This gives the bound,
\be
\label{pulsCcons}
C<1.6\times 10^{-9}~~~~~~~~~~~\text{(solitary pulsars)}\;.
\ee

\section{Conclusions}
\label{sec:conclusions}
We have constructed a linearized supergravity theory where
Lorentz invariance is broken down to
the subgroup of spatial rotations by a VEV of a timelike vector
field. This provides a supersymmetric extension of the well-known
Einstein-aether model. Our construction is based on embedding aether
as the lowest component into a chiral vector superfield which ensures
that aether VEV does not break SUSY. Using both the superfield
formalism and the ``on-shell'' component-field approach, we showed
that the linearized action contains, in addition to the usual SUGRA
parameters, a single free dimensionless coupling. This is to be
contrasted with the non-supersymmetric Einstein-aether model
possessing four arbitrary couplings.

We have derived the Lagrangian in terms of the physical component fields
and analyzed the spectrum of the theory. Due to
breaking of Lorentz invariance, the excitations with helicity 2 (graviton)
and 3/2 (gravitino) acquire the propagation velocity exceeding the
speed of light. We showed that this leads to the extension of the
on-shell gravity multiplet by two additional states with helicities
3/2 and 1. The extra states enter the theory as part of the aether
superfield and have the same superluminal velicity as graviton and
gravitino. The theory also contains one more pair of
helicity $\pm 1$ modes, two pairs of helicity
$\pm 1/2$ modes and two helicity 0 modes,
all propagating at the speed of light.
It is worth stressing that presence of superluminal
modes
does not lead to physical inconsistencies
as the theory features a preferred reference frame defined by the
aether VEV (cf. \cite{Jacobson:2008aj,Foster:2005dk}).

At low energies, upon SUSY breaking, the phenomenology of the model
reduces to that of the Einstein-aether theory with three out of the
four
couplings equal to zero. The strongest constraints on the model come
from the observation of the gravitational wave signal from the neutron
star merger GW170817 together with its electromagnetic counterpart
that limits the deviation of the gravity wave velocity from unity. It
can be translated into an upper bound on the energy scale of Lorentz
violations $M_*\lesssim 10^{11}$\,GeV. Independent, though weaker
constraints are imposed by the dynamics of the Solar System and
pulsars.

A natural development of our work will be its generalization to the
full non-linear supergravity case. This will open the way to study
possible manifestations of the super-aether model in cosmology, in particular, the
effects of the additional fermionic and bosonic fields present in the
model on the dynamics of the early universe. We plan to address this
topic in future.
As another direction, it would be
interesting to investigate applications of the model to the
holographic description of strongly coupled non-relativistic
systems.

Similarly to the non-supersymmetric Einstein-aether theory, the model
presented in this paper is a valid effective theory below the scale
$M_*$. One may wonder if and how the model can be UV completed above
this scale. In particular, if the completion can be achieved along the
lines of Ho\v rava gravity \cite{Horava:2009uw}. We do
not know the answer to this question and only note that a
supersymmetric extension of Ho\v rava gravity would need to overcome a
number of important obstructions discussed in \cite{Pujolas:2011sk}.

\paragraph{Acknowledgments} We are grateful to Luis Alvarez-Gaume,
Oriol Pujolas and Valery Rubakov for discussions. We thank Diego Blas,
Mikhail Ivanov and David Pirtskhalava
for valuable comments on the draft.
A.M. has been
supported by the Russian Foundation for Basic Research. The work of
S.S. was
supported in part by the Swiss National Science Foundation.
Research at the Perimeter Institute is supported in part by the
Government of Canada through the Department of Innovation, Science and
Economic Development Canada and
by the Province of Ontario
through the Ministry of Colleges and Universities.

\appendix

\section{Global SUSY transformations in components}
\label{app:globalSUSY}
Global SUSY acts on a general superfield $\Psi$ by translations in the
superspace with a co\-or\-di\-nate-independent spinor parameter $\zeta^\a$,
\[
\delta_G\Psi=(\zeta^\a Q_\a+\bar\zeta_{\dot\a}\bar Q^{\dot\a})\Psi\;,
\]
where the supercharges $Q_\a$, $\bar Q_{\dot\a}$ are realized as
differential operators on the superspace \cite{WB}.
Let a component field $\uppsi$ be
defined as
\[
\uppsi={\cal O}\Psi\big|\;,
\]
where ${\cal O}$ is an operator constructed of covariant derivatives $D_\a$, $\bar
D_{\dot\a}$.
Then its transformation equals to,
\[
\delta_G\uppsi={\cal O}(\zeta Q+\bar\zeta \bar Q)\Psi\big|
=(\zeta Q+\bar\zeta \bar Q){\cal O}\Psi\big|
=(\zeta D+\bar\zeta \bar D){\cal O}\Psi\big|\;,
\]
where the second equality holds because the supercharges anti-commute
with the covariant derivatives and the last equality holds because
the difference between
$Q_\a$ and $D_\a$ vanishes at zero $\theta$, $\bar\theta$.
Applying this formula to the component fields defined in
Eqs.~(\ref{Hcomp}), (\ref{compGamma}), (\ref{Vcomp}), we obtain:\\
For the gravitational supermultiplet,
\bseq
\label{dGH}
\begin{align}
\label{dGc}
\delta_Gc_m=&\zeta\chi_m+\bar\zeta\bar\chi_m\;,\\
\label{dGchi}
\delta_G\chi_{\a m}=&2\zeta_\a a_m-(\sigma^n\bar\zeta)_\a e_{nm}
+i(\sigma^n\bar\zeta)_\a\d_n c_m\;,\\
\label{dGa}
\delta_G a_m=&-\bar\zeta \bar\sigma_n\sigma_m\bar\psi^n
+i\bar\zeta\bar\sigma^n\d_n\chi_m\;,\\
\delta_Ge_{mn}=&-i\zeta\d_m\chi_n+i\bar\zeta\d_m\bar\chi_n
+i\zeta\sigma_m\bar\psi_n+i\zeta\sigma_n\bar\psi_m
+i\bar\zeta\bar\sigma_m\psi_n+i\bar\zeta\bar\sigma_n\psi_m\notag \\
&-i\upeta_{mn}(\zeta\sigma_k\bar\psi^k+\bar\zeta\bar\sigma_k\psi^k)
+\epsilon_{mnkl}(\zeta\sigma^k\bar\psi^l
-\bar\zeta\bar\sigma^k\psi^l)\;,
\label{dGe}\\
\label{dGpsi}
\delta_G\psi^m_\a=&\frac{i}{4}(\sigma^n\bar\sigma^m\zeta)_\a\,\Box c_n
+\frac{1}{2} (\sigma^l\bar\sigma^m\sigma^{nk}\zeta)_\a\,\d_k e_{nl}
-\frac{i}{2} (\sigma^n\bar\sigma^m\zeta)_\a\,d_n\;,\\
\label{dGd}
\delta_G
d_m=&\frac{1}{2}\zeta\Box\chi_m+\frac{1}{2}\bar\zeta\Box\bar\chi_m
+\frac{1}{2}\zeta\sigma_n\bar\sigma_k\sigma_m\d^n\bar\psi^k
-\frac{1}{2}\bar\zeta\bar\sigma_n\sigma_k\bar\sigma_m\d^n\psi^k\;;
\end{align}
\eseq
For the compensator supermultiplet,
\bseq
\label{dGGamma}
\begin{align}
\label{dGgamma}
\delta_G\gamma=&\zeta\phi+\bar\zeta\bar\omega\;,\\
\label{dGomega}
\delta_G\bar\omega_{\dot\a}=&(\zeta\sigma^m)_{\dot\a}\,q_m
-i(\zeta\sigma^m)_{\dot\a}\,\d_m\gamma\;,\\
\label{dGphi}
\delta_G\phi_\a=&2\zeta_\a B+(\sigma^m\bar\zeta)_\a\,q_m
+i(\sigma^m\bar\zeta)_\a\,\d_m\gamma\;,\\
\label{dGB}
\delta_G B=&-\bar\zeta\bar\nu+i\bar\zeta\bar\sigma^n\d_n\phi\;,\\
\label{dGq}
\delta_G q_m=&i\zeta\d_m\phi-i\bar\zeta\d_m\bar\omega
-i\bar\zeta\bar\sigma_m\sigma_k\d^k\bar\omega
+\zeta\sigma_m\bar\nu\;,\\
\label{dGnu}
\delta_G\bar\nu_{\dot\a}=&-i(\bar\zeta\bar\sigma^k\sigma^m)_{\dot\a}\,\d_kq_m
+\bar\zeta_{\dot\a}\,\Box\gamma\;;
\end{align}
\eseq
For the aether supermultiplet,
\bseq
\label{dGV}
\begin{align}
\label{dGv}
&\delta_G v_b=\zeta\eta_b+2iw^c\zeta\sigma_m\bar\sigma_{cb}\bar\psi^m
-w^c\zeta\sigma_{cb}\omega+2iw^c\bar\zeta\bar\sigma_m\sigma_{cb}\psi^m
-w^c\bar\zeta\bar\sigma_{cb}\bar\omega\;,\\
&\delta_G\eta_{b\a}=2i(\sigma^m\bar\zeta)_\a\,\d_mv_b\!+\!2\zeta_a\,G_b
\!-\!2iw^c (\sigma^{nm}\sigma_{cb}\sigma^k\bar\zeta)_\a\d_ne_{mk}
\!-\!2iw^c
(\sigma^k\bar\sigma_{cb}\bar\sigma^{nm}\bar\zeta)_\a\d_ne_{mk}\notag\\
&\qquad+w^c (\sigma_{cb}\sigma^m\bar\zeta)_\a (\Box c_m\!-\!2d_m\!+\!i\d_m\bar\gamma\!+\!\bar q_m)
+w^c (\sigma^m\bar\sigma_{cb}\bar\zeta)_\a(\Box c_m\!-\!2
d_m\!-\!i\d_m\gamma\!+\!q_m)\,,
\label{dGeta}\\
&\delta_G G_b=i\bar\zeta\bar\sigma^m\d_m\eta_b
-2w^c\bar\zeta\d_b\bar\psi_c+2w^c\bar\zeta\d_c\bar\psi_b
+2iw^c\epsilon_{cbmn}\bar\zeta\d^m\bar\psi^n
-2w^c\bar\zeta\bar\sigma^m\sigma^n\bar\sigma_{cb}\d_m\bar\psi_n\notag\\
&\qquad~~~~-iw^c\bar\zeta\bar\sigma^m\sigma_{cb}\d_m\omega+w^c\bar\zeta\bar\sigma_{cb}\bar\nu\;.
\label{dGG}
\end{align}
\eseq
Note that the transformations of the super-aether components depend on
the supergravity fields. This is a consequence of the dependence of
the covariant chirality
constraint (\ref{Vchir}) and the definition of aetherino (\ref{Vcomp})
on the superspace connection.

One observes that the above transformations in general violate the
gauge conditions (\ref{WZ}), (\ref{compfix}), (\ref{sime}). To restore
the gauge, they must be supplemented by appropriate gauge
transformations with the parameters depending on $\zeta^\a$ and the
fields. This leads to modified global SUSY transformations which we
denote with~$\tilde\delta_G$.

Let us find the parameters of the restoring gauge
transformations. Comparing the first three of Eqs.~(\ref{dGH})
with (\ref{comptrans1}) we see that to preserve the
Wess--Zumino gauge we have to choose,
\bseq
\label{gaugeG}
\be
\label{ximukapG}
\xi^I_m=0~,\qquad\mu_{m\a}=-2i(\sigma^n\bar\zeta)_\a\,e_{nm}~,\qquad
\kappa_m=2\bar\zeta\bar\sigma_n\sigma_m\bar\psi^n\;.
\ee
Next, preserving the gauge conditions (\ref{compfix}) requires,
\begin{align}
\label{trlambdaG}
&\lambda^\a_{~\a}=\frac{(3n\!+\!1)(n\!+\!1)}{4n}(\zeta\phi\!+\!\bar\zeta\bar\omega)
+\frac{(3n\!+\!1)(n\!-\!1)}{4n}(\bar\zeta\bar\phi\!+\!\zeta\omega)
+i(n\!+\!1)(\zeta\sigma_k\bar\psi^k\!+\!\bar\zeta\bar\sigma_k\psi^k)\,,\\
\label{rhoG}
&\bar\rho^{\dot\a}=\frac{i}{2}(\bar\sigma^m\zeta)^{\dot\a}\d^ke_{km}
\!-\!\frac{i}{2}(\bar\sigma^m\zeta)^{\dot\a}\d_me^k_{~k}
-n(\bar\sigma^m\zeta)^{\dot\a}d_m+\frac{3n\!+\!1}{2}(\bar\sigma^m\zeta)^{\dot\a}q_m^R
-\frac{3n\!+\!1}{2}\bar\zeta^{\dot\a}\bar B\,.
\end{align}
Finally, the symmetry of the tetrad is preserved if we choose,
\be
\label{notrlambdaG}
(\sigma_{mn})^{\a\b}\lambda_{\a\b}
-(\bar\sigma_{mn})^{\dot\a\dot\b}\bar\lambda_{\dot\a\dot\b}
=-\epsilon_{mnkl}(\zeta\sigma^k\bar\psi^l-\bar\zeta\bar\sigma^k\psi^l)\;.
\ee
\eseq
The remaining parameters $\xi^R_m$ and $\varepsilon^\a$ do not need to
be adjusted and describe the unconstrained gauge symmetries.

The modified global SUSY transformations are obtained by adding to
(\ref{dGH}), (\ref{dGGamma}), (\ref{dGV}) the gauge shifts
(\ref{comptrans}), (\ref{lowGtrans}), (\ref{compVtrans}) with the
parameters given by (\ref{gaugeG}). We work them out only
for the physical fields $e_{mn}$, $\psi^m_\a$, $v_b$ and
$\eta_{b\a}$. Using the expressions (\ref{trlambdaG}),
(\ref{notrlambdaG}) we find for the tetrad,
\[
\tilde\delta_G e_{mn}=i\zeta\s_m\bar\psi_n+i\zeta\s_n\bar\psi_m
+\frac{3n+1}{4}\upeta_{mn}\,\zeta\Big(\omega+\phi+\frac{4in}{3n+1}\s_k\bar\psi^k\Big)
+\text{h.c.}
\]
Notice that the linear combination in brackets is the same as in the
gauge condition (\ref{superconffix}). Setting it to zero we eliminate
the term proportional to $\upeta_{mn}$ and arrive to the canonical
SUSY transformation of the metric (\ref{tildedGh}).
For gravitino, we use Eq.~(\ref{rhoG}) and the symmetry of the
tertrad to simplify the expression. This yields,
\be
\label{tildedGpsi}
\begin{split}
\tilde\delta_G\psi^m_\a=\;&\frac{1}{2}(\s^{kn}\zeta)_\a\d_kh_{n}^{~m}
+i\zeta_\a d_m+\frac{i(n\!+\!1)}{2}(\s^m\bar\s^n\zeta)_\a \bigg[d_n
-\frac{3n\!+\!1}{2(n\!+\!1)} q_m^R\bigg]\\
&+\frac{i(3n\!+\!1)}{4}(\s^m\bar\zeta)_\a\bar B\,.
\end{split}
\ee
Finally, for aether and aetherino upon using the gauge conditions and
some simplifications we obtain,
\bseq
\label{tildedGV}
\begin{align}
\label{tildedGv}
&\tilde\delta_G v_b=\zeta\eta_b
+iw^c(\zeta\s_c\bar\psi_b-\zeta\s_b\bar\psi_c+\bar\zeta\bar\s_c\psi_b
-\bar\zeta\bar\s_b\psi_c)
-w^c(\zeta\s_{cb}\omega+\bar\zeta\bar\s_{cb}\bar\omega)\;,\\
&\tilde\delta_G\eta_{b\a}=2i(\s^m\bar\zeta)_\a\d_mv_b+2\zeta_\a G_b
+iw^c (\s^k\bar\zeta)_\a (\d_ch_{kb}-\d_bh_{kc})\notag\\
&\quad-\frac{i}{2}w^c(\s_c\bar\zeta)_\a\d^kh_{kb}
\!+\frac{i}{2}w^c(\s_b\bar\zeta)_\a\d^kh_{kc}
\!+\!\frac{in}{3n\!+\!1}w^c(\s_c\bar\zeta)_\a\d_bh
\!-\!\frac{in}{3n\!+\!1}w^c(\s_b\bar\zeta)_\a\d_ch\notag\\
&\quad+iw^c\epsilon_{cbmn}(\s^m\bar\zeta)_\a(2d^n-q^{R,n})
+iw^c(\s_c\bar\zeta)_\a q_b^I-iw^c(\s_b\bar\zeta)_\a q_c^I\;.
\label{tildedGeta}
\end{align}
\eseq
Notice that the additional terms containing the gravitational fields
in (\ref{tildedGv}) are purely real and thus contribute only to the
transformation of $v^R_b$. The expressions (\ref{tildedGpsi}),
(\ref{tildedGV})
are used in Sec.~\ref{sec:fermionic} to derive the ``on-shell'' SUSY
algebra.

\section{Torsion constraints in linearized SUGRA}
\label{app:torsion}

In this Appendix we linearize the constraints on the torsion in
superspace and use them to derive the expressions for the
superspace connection in terms of the fields of linear non-minimal
SUGRA. At the linearized level the torsion tensor is related to the
connection and vielbein as follows \cite{Girardi:1984eq},
\be
\label{torlin}
\begin{split}
T_{CB}^{~~~A}=&T_{CB}^{(0)~A}+\Phi_{CB}^{~~~A}-(-1)^{|B||C|}\Phi_{BC}^{~~~A}
+D_CI_B^{~\;A}-(-1)^{|B||C|}D_BI_C^{~\;A}\\
&+T_{CB}^{(0)~D}I_D^{~\;A}-I_C^{~\;D}T_{DB}^{(0)~A}
+(-1)^{|B||C|}I_B^{~\;D}T_{DC}^{(0)~A}\;,
\end{split}
\ee
where the tensor $I_B^{~\;A}$ describing fluctuations of the
vielbein has been defined in (\ref{viellin}) and
$T_{CB}^{(0)~A}$ is the flat-superspace torsion whose only
non-vanishing components are
$T_{\g\dot\b}^{(0)~a}=T_{\dot\b\g}^{(0)~a}=2i\s^a_{\g\dot\b}$\,. The torsion (\ref{torlin})
satisfies the following constraints \cite{Girardi:1984eq},
\bseq
\label{torcons}
\begin{gather}
\label{torcons1}
T_{\g\b}^{\phantom{\g\b}a}=T^{\dot\g\dot\b a}
=T_{\g\b\dot\a}=T^{\dot\g\dot\b\a}=0\;,\\
\label{torcons2}
T_{\g\dot\b}^{\phantom{\g\b}a}=T_{\dot\b\g}^{\phantom{\g\b}a}=2i\s_{\g\dot\b}^a\;,\\
\label{torcons3}
T^{~\dot\b}_{\g~\dot\a}=(n-1)\delta_{\dot\a}^{\dot\b}\,T_\g\;,~~~~
T_{~\b}^{\dot\g~\a}=(n-1)\delta_{\b}^{\a}\,\bar T^{\dot\g}\;,\\
\label{torcons4}
T_{\g\b}^{\phantom{\g\b}\a}=(n+1)(\delta_\g^\a\,T_\b+\delta^\a_\b\,T_\g)\;,~~~~
T^{\dot\g\dot\b}_{\phantom{\g\b}\dot\a}=
(n+1)(\delta^{\dot\g}_{\dot\a}\,\bar
T^{\dot\b}+\delta_{\dot\a}^{\dot\b}\,\bar T^{\dot\g})\;,\\
\label{torcons5}
T_{\g b}^{\phantom{\g b}a}=2n\delta_b^a\,T_\g\;,~~~~
T_{~b}^{\dot\g~a}=2n\delta_b^a\,\bar T^{\dot\g}\;,\\
\label{torcons6}
T_{cb}^{\phantom{cb}a}=0\;,
\end{gather}
\eseq
where the Bianchi identities imply that the superfield $T_\g$
and its conjugate obey
\be
\label{Tcons}
D_\a T_\g+D_\g T_\a=0~,~~~
\bar D_{\dot\a}\bar T_{\dot\g}+\bar D_{\dot\g} \bar T_{\dot\a}=0\;.
\ee
Our strategy is to apply the constraints (\ref{torcons}) to the
relation (\ref{torlin}) and, using the components of $I_B^{~\;A}$
found in Sec.~\ref{sec:2.2}, derive the equations for the remaining
vielbein components and connection.

The first set of constraints (\ref{torcons1}) are trivially satisfied
by the vielbein (\ref{viel1}) and do not provide any further
information. Inserting (\ref{viel1}) into
(\ref{torcons2}) we read off the components
\be
\label{viel2}
I_a^{~\;b}=-\delta_a^b\,\frac{1}{2}(\Gamma'+\bar\Gamma')
-\Delta_a H^b\;.
\ee
From (\ref{torcons3}) and (\ref{torcons5}) we get respectively
\bseq
\label{torint}
\begin{align}
\label{torint1}
&(n-1)\delta^\a_\b\,\bar T_{\dot\g}=\Phi_{\dot\g\b}^{\phantom{\g\b}\a}
+\bar D_{\dot\g}I_\b^{~\a}+2i I_{\b\dot\g}^{\phantom{\b\g}\a}\;,\\
\label{torint2}
&2n\delta^a_b\,\bar T_{\dot\g}=\Phi_{\dot\g b}^{\phantom{\g b} a}
+\bar D_{\dot\g}I_b^{~a}-\d_bI_{\dot\g}^{~a}-i\s_{\delta\dot\g}^a
\bar\s^{\dot\b\b}_b\,I_{\b\dot\b}^{\phantom{\b\b}\delta}\;,
\end{align}
\eseq
where we have introduced
$I_{\b\dot\b}^{\phantom{\b\b}\delta}\equiv\s_{\b\dot\b}^b
I_b^{~\delta}$. Let us take the trace of these equations. This
eliminates the connection, which is traceless, and one is left with,
\bseq
\label{tortrace}
\begin{align}
\label{tortrace1}
&2(n-1)\,\bar T_{\dot\g}=\bar D_{\dot\g}I_\b^{~\b}
+2i I_{\b\dot\g}^{\phantom{\b\g}\b}\;,\\
\label{tortrace2}
&8n\,\bar T_{\dot\g}=\bar D_{\dot\g}I_b^{~b}-\d_bI_{\dot\g}^{~b}
+2iI_{\b\dot\g}^{\phantom{\b\b}\b}\;.
\end{align}
\eseq
This system can be solved for the two unknowns $\bar T_{\dot\g}$,
$I_{\b\dot\g}^{\phantom{\b\b}\b}$. Using the expressions
(\ref{viel1}), (\ref{viel2}) we obtain,
\bseq
\label{IbgbTg}
\begin{align}
\label{Ibgb}
&I_{\b\dot\g}^{\phantom{\b\b}\b}=i\bar D_{\dot\g}\Gamma
+\frac{i}{8}\bar D^2D^\b H_{\b\dot\g}\;,\\
\label{Tg}
&\bar T_{\dot\g}=\frac{3n+1}{8n}\bar D_{\dot\g}\bar\Gamma
+\frac{3n-1}{8n}\bar D_{\dot\g}\Gamma
+\frac{n-1}{32n}\bar D^2D^\a H_{\a\dot\g}
-\frac{n+1}{16n}\bar D_{\dot\g}D^\a\bar D^{\dot\a}H_{\a\dot\a}\;.
\end{align}
\eseq
Note that the expression for $\bar T_{\dot\g}$ satisfies the condition
(\ref{Tcons}). We now use the remaining information contained in
Eqs.~(\ref{torint}). We take the symmetric part of
Eq.~(\ref{torint1}),
\bseq
\label{torsymm}
\be
\label{torsymm1}
2\Phi_{\dot\g\a\b}+2i(I_{\a\dot\g\b}+I_{\b\dot\g\a})=0\;,
\ee
where we have used the symmetry of the connection in the last two
indices. Multiplying Eq.~(\ref{torint2}) by $(\s_{ba})_{\a\b}$ and
$(\bar\s_{ba})_{\dot\a\dot\b}$ and using the constitutive relation
(\ref{Phistruct}) we obtain,
\begin{align}
&-2\Phi_{\dot\g\a\b}-i(I_{\a\dot\g\b}+I_{\b\dot\g\a})=
-(\s_{ba})_{\a\b}\bar D_{\dot\g}I_{ba}+(\s_{ba})_{\a\b}\d_bI_{\dot\g
  a}\;,\\
&-2\Phi_{\dot\g\dot\a\dot\b}=-(\bar\s_{ba})_{\dot\a\dot\b}\bar
D_{\dot\g}I_{ba}
+(\bar\s_{ba})_{\dot\a\dot\b}\d_bI_{\dot\g a}
-i\epsilon_{\dot\a\dot\g}I_{\g\dot\b}^{\phantom{\g\dot\b}\g}
-i\epsilon_{\dot\b\dot\g}I_{\g\dot\a}^{\phantom{\g\dot\a}\g}\;,
\end{align}
\eseq
where we put on the r.h.s. the components that are already
known. This system allows us to find the connection components
$\Phi_{\dot\g\a\b}$, $\Phi_{\dot\g\dot\a\dot\b}$ and the symmetrized
perturbation of the vielbein $I_{\a\dot\g\b}+I_{\b\dot\g\a}$. The
results for the connections are given in Eqs.~(\ref{connspin}) of the
main text, whereas for the vielbein adding the trace part (\ref{Ibgb})
we obtain,
\be
\label{viel3}
I_{\a\dot\a}^{\phantom{\a\a}\b}=
\frac{i}{2}\delta_\a^\b\bar D_{\dot\a}\Gamma
-\frac{i}{8}\bar D^2D_\a H^\b_{~\;\dot\a}\;.
\ee
This completes the determination of the linearized vielbein.
The connection $\Phi_{\dot\g ab}$ follows from the constitutive
relation (\ref{Phistruct}) and is given in Eq.~(\ref{connvect}).
Alternatively, it can be found from Eq.~(\ref{torint2})
using the expressions for the vielbein. One can check with the
formulas derived above that the constraints (\ref{torcons4}) are
automatically satisfied. Finally, the constraint (\ref{torcons6})
provides an equation for the connection components
$\Phi_{cb}^{\phantom{cb}a}$, which we do not use in this paper.

\section{Calculus in superspace}
\label{app:useful}

\subsection{Relations between superfield operators}
\label{app:Fierz}

The number of independent operators that can appear in the superfield
Lagrangian for linearized SUGRA with broken Lorentz symmetry is
reduced by various relations between them arising as a consequence of spinor
algebra. The following properties of the superspace differential
operators are used in the calculation:\\
commutators:
\be
\label{Dcommut}
[\bar D_{\dot\a},D^2]=4i\d_{\g\dot\a}D^\g~,~~~~
[\bar D^2,D_\b]=4i\d_{\b\dot\g}\bar D^{\dot\g}~,~~~~
[\DD_a,\DD_b]=2\epsilon_{abmn}\d^m\DD^n\;;
\ee
rules for integration by parts:
\be
\label{intrules}
\Psi_1 D^2\Psi_2\simeq (D^2\Psi_1) \Psi_2~,~~~~
\Psi_1 \bar D^2\Psi_2\simeq (\bar D^2\Psi_1) \Psi_2~,~~~~
\Psi_1 \DD\Psi_2\simeq (\DD\Psi_1) \Psi_2\;,
\ee
where $\Psi_{1,2}$ are arbitrary superfields and the sign $\simeq$
stands for equality up to a total derivative. Using these relations
one derives the identities,
\bseq
\label{Fierz}
\begin{align}
\label{Fierz1}
\d_a H^{(b}\DD_c H^{d)}\simeq&\; 0\;,\\
\label{Fierz1a}
\d_a H^{(b} D^2 H^{d)}\simeq&\;\DD_a H^{(b} D^2 H^{d)}\simeq\; 0\;,\\
\label{Fierz2}
D^2H^{(b}\bar D^2 H^{d)}\simeq&-2\,\DD_mH^b\DD^mH^d-6\,\d_mH^b\d^mH^d\;,\\
\label{Fierz3}
\DD_{(a} H^b\DD_{c)}H^d\simeq&\;
\d_{(a} H^b\d_{c)}H^d
+\frac{1}{4}\eta_{ac}(\DD_m H^b\DD^m H^d-\d_mH^b\d^mH^d)\;,\\
\label{Fierz4}
\DD_aH_k\DD_c H^k\simeq&\;\d_a H_k\d_cH^k
+\frac{1}{4}\eta_{ac}(\DD_m H_k\DD^m H^k-\d_mH_k\d^mH^k)\;,\\
\label{Fierz5}
\DD_a H^k\DD_k H_b\simeq& -\DD_a H_b\DD_k H^k
+2\d_a H_b\d_kH^k
+\frac{1}{2}(\DD_mH_a\DD^m H_b-\d_mH_a\d^mH_b)\;,\\
\label{Fierz6}
\DD_a H^{[c}\DD_b H^{d]}\simeq& -\epsilon_{abmn}\d_m H^{[c}\DD_n
H^{d]}\;,\\
\label{Fierz7}
\DD_{[a} H^{c}\DD_{b]} H^{d}\simeq& -\epsilon_{abmn}\d_m H^{c}\DD_n
H^{d}\;,\\
\label{Fierz8}
\epsilon_{amnk}\d^m H_b\DD^nH^k\simeq& -\DD_aH_b\DD_kH^k
+\d_aH_b\d_kH^k
+\frac{1}{4}(\DD_mH_a\DD^mH^b-\d_mH_a\d^m H_b)\\
\epsilon_{amnk}\d_bH^m\DD^n H^k\simeq&
-\epsilon_{amnk}\DD_bH^m\d^n H^k
-2\DD_a H_b\DD_kH^k+2\d_a H_b\d_kH^k\notag\\
&+\frac{1}{2}(\DD_kH_a\DD^kH_b-\d_kH_a\d^kH_b)\notag\\
&+\eta_{ab}\bigg((\DD_kH^k)^2-(\d_kH^k)^2
-\frac{1}{4}\DD_mH_k\DD^mH^k+\frac{1}{4}\d_mH_k\d^mH^k\bigg)
\;.
\label{Fierz9}
\end{align}
\eseq
Here the round (square) brackets denote symmetrization (antisymmetrization)
over the corresponding
indices.

\subsection{Operators without aether perturbation}
\label{app:classify}

It is convenient to further subdivide these terms 
according to the number of
insertions of the spurion $w^a$. It is straightforward to see that
the maximal number of
insertions is 4. Thus, we have:

{\it 4 insertions of $w^a$.} There is a single independent operator,
\be
\label{4w}
w^a w^b w^c w^d \d_a H_b \d_cH_d\;.
\ee
Two other possible operators would be
\[
w^a w^b w^c w^d \d_a H_b \DD_cH_d\;,~~~~~
w^a w^b w^c w^d \Delta_a H_b \Delta_cH_d\;.
\]
However, the first of them is a total derivative, see
Eq.~(\ref{Fierz1}) in Appendix~\ref{app:Fierz},
whereas the second is expressed in terms of (\ref{4w}) and
contributions with fewer insertions of $w^a$ due to the relation
(\ref{Fierz3}).

{\it 3 insertions of $w^a$.} This group is actually empty. The
operators that can be written using three $w^a$-insertions are
\[
w^a w^b w^c \d_aH_b D^2 H_c~,~~~~w^a w^b w^c \DD_aH_b D^2 H_c
\]
and their complex conjugate. However, they vanish upon integration over
the superspace, see Eq.~(\ref{Fierz1a}).

{\it 2 insertions of $w^a$}. There are in total 12 independent operators
that we choose as follows,
\vspace{-0.5cm}
\bseq
\label{2w}
\begin{gather}
\label{2w1}
w^a w^b \d_aH_b \d_c H^c~,~~~~w^a w^b \d_aH_c \d_b H^c~,~~~~
w^a w^b \d_cH_a \d^c H_b\;,\\
\label{2w2}
w^a w^b \DD_aH_b \DD_c H^c~,~~~~
w^a w^b \DD_cH_a \DD^c H_b\;,\\
\label{2w3}
w^a w^b \d_aH_b \DD_c H^c~,~~~~
w^a w^b \d_cH^c \DD_a H_b\;,\\
\label{2w5}
w^a w^b \epsilon_{acde}\d^cH^d\DD_bH^e\;,\\
\label{2w7}
w^a w^b \d_aH_b\Gamma~,~~~~
w^a w^b \d_aH_b\bar\Gamma~,~~~~
w^a w^b \DD_aH_b\Gamma~,~~~~
w^a w^b \DD_aH_b\bar\Gamma\;.
\end{gather}
\eseq
Other operators that can be written using two $w^a$ insertions are
\bseq
\label{2wextra}
\begin{gather}
\label{2wextra2}
w^a w^b \DD_aH_c \DD_b H^c~,~~~~w^a w^b \DD_aH_c \DD^c H_b\;,\\
\label{2wextra3}
w^a w^b \d_aH_c \DD_b H^c~,~~~~w^a w^b \d_cH_a \DD^c H_b~,~~~~
w^a w^b \epsilon_{acde}\d_bH^c\d^dH^e\;,\\
\label{2wextra4}
w^a w^b \epsilon_{acde}\DD_bH^c\DD^dH^e~,~~~~
w^a w^b \epsilon_{acde}\DD^cH_b\DD^dH^e\;,\\
\label{2wextra5}
w^a w^b \epsilon_{acde}\d^cH_b\DD^dH^e~,~~~~
w^a w^b \epsilon_{acde}\d_bH^c\DD^dH^e\;,\\
\label{2wextra6}
w^a w^b D^2H_a \bar D^2 H_b\;.
\end{gather}
\eseq
Using the identities (\ref{Fierz}) one shows
that the contributions of the latter operators into the Lagrangian are
degenerate with the
operators (\ref{2w}): for the two operators (\ref{2wextra2}) this is
due to the relations (\ref{Fierz4}) and (\ref{Fierz5});
the terms (\ref{2wextra3})
vanish upon integration; the operators
(\ref{2wextra4}) and (\ref{2wextra5})
are eliminated using (\ref{Fierz6}) --- (\ref{Fierz9});
the operator (\ref{2wextra6}) is eliminated due to (\ref{Fierz2}).

{\it 1 insertion of $w^a$.} There are 2 terms,
\be
\label{1w}
w^a \d_b H^b D^2 H_a~,~~~~
w^a D^2 H_a \Gamma\;,
\ee
and their complex conjugate. One more operator
\[
w^a \d_a H_b D^2 H^b
\]
is a total derivative, see (\ref{Fierz1a}). Also, a replacement of the
ordinary derivative
$\d_a$ by $\DD_a$ in the above operators
does not generate new
contributions due to the anti-chirality of the field~$D^2H_a$.

{\it No insertions of $w^a$.} 12 independent operators are,
\bseq
\label{0w}
\begin{gather}
\label{0w1}
(\d_a H^a)^2~,~~~~ \d_aH_b \d^a H^b\;,\\
\label{0w2}
(\DD_a H^a)^2~,~~~~ \DD_aH_b \DD^a H^b\;,\\
\label{0w3}
\d_a H^a \DD_b H^b\;,\\
\label{0w5}
\d_a H^a\Gamma~,~~~~\d_a H^a\bar\Gamma~,~~~~
\DD_a H^a\Gamma~,~~~~\DD_a H^a\bar\Gamma\;,\\
\label{0w6}
\Gamma^2~,~~~~ \bar\Gamma^2~,~~~~ \Gamma\bar\Gamma\;.
\end{gather}
\eseq
The five remaining combinations,
\be
\label{Owextra}
\d_aH_b \DD^a H^b\,,~~~~
D^2H_a\bar D^2H^a\,,~~~~\DD_aH_b\DD^b H^a\,,~~~~
\epsilon_{abcd}\DD^aH^b\DD^c H^d\,,~~~~
\epsilon_{abcd}\d^aH^b\DD^c H^d\,,
\ee
produce degenerate contributions into the action, as it follows from
Eqs.~(\ref{Fierz1}), (\ref{Fierz2}), (\ref{Fierz5}), (\ref{Fierz6}),
(\ref{Fierz8}).

\subsection{Transformations of the superfield operators}
\label{app:trans}
In this Appendix we derive the gauge variations of various superfield
operators 
that can potentially enter into the quadratic Lagrangian. The
operators (\ref{quadrV}), (\ref{linV}) have been discussed in the main
text. Here we focus on the operators from
Appendix~\ref{app:classify}. 

We start with the operator (\ref{4w}). Its variation reads,
\be
\label{4wtrans}
\delta_L (w^aw^bw^cw^d\d_aH_b\d_cH_d)\simeq
L_\b w^aw^bw^cw^d\bar\s_a^{\dot\b\b}\bar D_{\dot\b}\d_b\d_c H_d+\text{h.c.}\;,
\ee
and manifestly contains four insertions of $w^a$. There are no other
operators whose variation would have this property and therefore
(\ref{4w})
is also absent from
the invariant action.

To proceed, we notice that the remaining operators split into several
sectors which do not mix under the linearized super-diffeomorphisms. These
sectors are characterized by the properties of the operators under the
action of the $R$-symmetry and $CP$. The $R$-symmetry rotates the
phases of the spinor derivatives,
\be
\label{Rsym}
D_\a\mapsto\e^{-i\varphi}D_\a~,~~~~~
\bar D_{\dot\a}\mapsto\e^{i\varphi}\bar D_{\dot\a}\;,
\ee
with the superfields $H_a$, $\Gamma$, $V^a$ and the ordinary
derivatives $\d_a$ kept intact. Correspondingly, the operators
(\ref{2w}), (\ref{0w}) have zero $R$-charge, whereas the $R$-charge of
the operators (\ref{1w}) is $-2$. The $R$-charge is preserved by the
super-diffeos, provided one assigns $R=-1$ to the gauge parameter
$L_\b$. This implies that if the operators (\ref{1w}) entered into the
invariant action, their variations would have to cancel with each
other. However, it is straightforward to see that this is
impossible. We omit the operators (\ref{1w}) in what follows.

Next we turn to the properties of the operators under parity. Pure
parity does not preserve the SUSY algebra and thus cannot be defined
on the superspace. To be compatible with SUSY, parity must be
supplemented by the charge conjugation \cite{Gates:1983nr}. In the
Lorentz frame where the spatial components of the
aether VEV vanish, see (\ref{wtimel}),
the $CP$ transformations have the
form\footnote{Note that in this frame $V_0=0$ due to the constraint
  (\ref{Vnorm}).},
\bseq
\begin{gather}
V_i\mapsto -\bar V_i\;\\
H_0\mapsto -H_0~,~~~~H_i\mapsto H_i~,~~~~\Gamma\mapsto\bar\Gamma\;\\
\d_0\mapsto \d_0~,~~~~\d_i\mapsto -\d_i\;,\\
\DD_0\mapsto -\DD_0~,~~~~\DD_i\mapsto \DD_i\;,
\end{gather}
\eseq
where $i=1,2,3$ denote the spatial indices. Notice that $(V_a+\bar V_a)$,
$\d_a$ transform as vectors, whereas $(V_a-\bar V_a)$, $H_a$, $\DD_a$ are
pseudo-vectors. Clearly, the SUGRA action (\ref{NMact}) is
$CP$-even. It is convenient to choose the basis of operators having
definite $CP$ quantum numbers. Out of (\ref{2w}), (\ref{0w}) we
construct the following combinations:

{\bf $CP$-even:}
\bseq
\label{CPeven}
\begin{gather}
w^aw^b\d_aH_b\d_cH^c~,~~~~w^aw^b\d_aH_c\d_bH^c~,~~~~w^aw^b\d_cH_a\d^cH_b\;,\\
w^aw^b\DD_aH_b\DD_cH^c~,~~~~w^aw^b\DD_cH_a\DD^cH_b~,~~~~
w^aw^b\epsilon_{acde}\d^cH^d\DD_bH^e\;,\\
iw^aw^b\d_aH_b(\Gamma-\bar\Gamma)~,~~~~
w^aw^b\DD_aH_b(\Gamma+\bar\Gamma)\;,\\
(\d_aH^a)^2~,~~~~\d_aH_b\d^a
H^b~,~~~~(\DD_aH^a)^2~,~~~~\DD_aH_b\DD^aH^b\;,\\
i\d_aH^a(\Gamma-\bar\Gamma)~,~~~\DD_aH^a(\Gamma+\bar\Gamma)~,~~~~
\Gamma^2+\bar\Gamma^2~,~~~~\Gamma\bar\Gamma\;;
\end{gather}
\eseq

{\bf $CP$-odd:}
\bseq
\label{CPodd}
\begin{gather}
w^aw^b\d_aH_b\DD_cH^c~,~~~~w^aw^b\DD_aH_b\d_cH^c\;,\\
w^aw^b\d_aH_b(\Gamma+\bar\Gamma)~,~~~~
iw^aw^b\DD_aH_b(\Gamma-\bar\Gamma)\;,\\
\d_aH^a\DD_bH^b~,~~~~
\d_aH^a(\Gamma+\bar\Gamma)~,~~~i\DD_aH^a(\Gamma-\bar\Gamma)~,~~~~
i(\Gamma^2-\bar\Gamma^2)\;.
\end{gather}
\eseq
To the first group one has to add the $CP$-even operator
(\ref{quadrV}). The gauge variations of operators should cancel
separately within each group.

We expand
the variations as linear combinations
of independent contributions. For the
$CP$-even sector the coefficients of this expansion are listed in
Table~\ref{tab:1} and for the $CP$-odd sector in
Table~\ref{tab:2}.
Each column in the tables
corresponds to a given operator and rows to the
terms in the expansion of its variation.
The notations for the rows are,
\bseq
\begin{align}
&{\cal O}_1=L_\b w^aw^b\bar\s_a^{\dot\b\b}\bar D_{\dot\b}\d_b\d_k
H^k\;,\\
&{\cal O}_2=L_\b w^aw^b\bar\s_k^{\dot\b\b}\bar D_{\dot\b}\d^k\d_a
H_b\;,\\
&{\cal O}_3=L_\b w^aw^b\bar\s_k^{\dot\b\b}\bar D_{\dot\b}\d_a\d_b
H^k\;,\\
&{\cal O}_4=L_\b w^aw^b\bar\s_a^{\dot\b\b}\bar D_{\dot\b}\Box H_b\;,\\
&{\cal O}_5=L_\b w^aw^b(\s_{ak})_\g^{~\;\b}\bar D^2D^\g\d_bH^k\;,\\
&{\cal O}_6=L_\b w^aw^b(\s_{ak})_\g^{~\;\b}\bar D^2D^\g\d^kH_b\;,\\
&{\cal O}_7=L_\b w^aw^b\bar D^2D^\b\d_a H_b\;,\\
&{\cal O}_8=L_\b w^aw^b\epsilon_{aklm}\bar\s^{k\dot\b\b}\bar D_{\dot\b}\d_b\d^l
H^m\;,\\
&{\cal O}_9=L_\b w^aw^b\bar\s_a^{\dot\b\b}\bar
D_{\dot\b}\d_b\Gamma\;,\\
&{\cal O}_{10}=L_\b w^aw^b\bar\s_a^{\dot\b\b}\bar
D_{\dot\b}\d_b\bar\Gamma\;,\\
&{\cal O}_{11}=L_\b (\s_{kl})_\g^{~\;\b}\bar D^2 D^\g\d^kH^l\;,\\
&{\cal O}_{12}=L_\b\bar\s_k^{\dot\b\b}\bar D_{\dot\b}\d^k\d_lH^l\;,\\
&{\cal O}_{13}=L_\b\bar\s_k^{\dot\b\b}\bar D_{\dot\b}\Box H^k\;,\\
&{\cal O}_{14}=L_\b\bar D^2 D^\b\d_k H^k\;,\\
&{\cal O}_{15}=L_\b\bar\s_k^{\dot\b\b}\bar D_{\dot\b}\d^k\Gamma\;,\\
&{\cal O}_{16}=L_\b\bar\s_k^{\dot\b\b}\bar
D_{\dot\b}\d^k\bar\Gamma\;,\\
&{\cal O}_{17}=L_\b\bar D^2 D^\b\bar\Gamma\;.
\end{align}
\eseq
We focus only on the part of the variations proportional to $L_\b$;
the terms with $\bar L_{\dot\b}$ are obtained by complex
conjugation. For example, from the second column of Table~\ref{tab:1}
one reads,
\[
\delta_L(w^aw^b\d_aH_b\d_cH^c)\simeq \frac{1}{2}{\cal
  O}_1+\frac{1}{2}{\cal O}_2+\text{h.c.}\;,
\]
where $\simeq$ stands, as usual, for `equal up to a total derivative'.

A gauge invarinat linear combination of operators with a vector of
coefficients $X$ corresponds to a solution of
the system of equations,
\be
\label{MX}
{\cal M}\cdot X=0\;,
\ee
where ${\cal M}$ is the matrix of Table~\ref{tab:1} (\ref{tab:2}) for
the $CP$-even (odd) sector respectively. In the case of $CP$-even
operators a non-trivial solution of this system exists and is
parameterized by two free variables. The corresponding invariant
action is presented in the main text. For the $CP$-odd case
Eq.~(\ref{MX}) has only a trivial solution, $X=0$.
\begin{landscape}
\pagestyle{empty}
\begin{table}
\centering
\begin{tabular}{|c|c|c|c|c|c|c|c|c|c|c|c|c|c|c|c|c|c|}
\hline
&
\begin{sideways}
$\delta_L(V_a\bar V^a)$
\end{sideways}
&\begin{sideways} $\delta_L(w^aw^b\d_aH_b\d_cH^c)$  \end{sideways}
&\begin{sideways} $\delta_L(w^aw^b\d_aH_c\d_bH^c)$  \end{sideways}
&\begin{sideways} $\delta_L(w^aw^b\d_cH_a\d^cH_b)$  \end{sideways}
&\begin{sideways} $\delta_L(w^aw^b\DD_aH_b\DD_cH^c)$  \end{sideways}
&\begin{sideways} $\delta_L(w^aw^b\DD_cH_a\DD^cH_b)$  \end{sideways}
&\begin{sideways} $\delta_L(w^aw^b\epsilon_{acde}\d^cH^d\DD_bH^e)$  \end{sideways}

&\begin{sideways} $\delta_L\big(iw^aw^b\d_aH_b(\Gamma-\bar\Gamma)\big)$  \end{sideways}
&\begin{sideways} $\delta_L\big(w^aw^b\DD_aH_b(\Gamma+\bar\Gamma)\big)$  \end{sideways}
&\begin{sideways} $\delta_L\big((\d_aH^a)^2\big)$  \end{sideways}
&\begin{sideways} $\delta_L(\d_aH_b\d^aH^b)$  \end{sideways}
&\begin{sideways} $\delta_L\big((\DD_aH^a)^2\big)$  \end{sideways}
&\begin{sideways} $\delta_L\big(\DD_aH_b\DD^aH^b\big)$  \end{sideways}
&\begin{sideways} $\delta_L\big(i\d_aH^a(\Gamma-\bar\Gamma)\big)$  \end{sideways}
&\begin{sideways}
  $\delta_L\big(\DD_aH^a(\Gamma+\bar\Gamma)\big)$  \end{sideways}
&\begin{sideways} $\delta_L(\Gamma^2+\bar\Gamma^2)$  \end{sideways}
&\begin{sideways} $\delta_L(\Gamma \bar\Gamma)$  \end{sideways}
\\
\hline
${\cal O}_1$ & &$\frac{1}{2}$ & & &$\frac{1}{2}$ & &  & & & & & & & &
& & \\
 \hline
${\cal O}_2$ & &$\frac{1}{2}$ & & &$\frac{1}{2}$ & & &
$\frac{n+1}{3n+1}$&$-\frac{n+1}{3n+1}$  & & & & & & & &\\
 \hline
${\cal O}_3$ & & &$1$ & & & &  & & & & & & & & & & \\
 \hline
${\cal O}_4$ & & & &$1$& &$1$ & & & & & & &  & & & &\\
 \hline
${\cal O}_5$ & & & & &$-\frac{i}{4}$ & & $-\frac{i}{2}$& & & & & &  & & & &\\
 \hline
${\cal O}_6$ &$-\frac{i}{2}$ & & & & $-\frac{3i}{4}$ &$-2i$
&$\frac{i}{2}$ & &$\frac{i}{2}$ & &  & & & & & &\\
 \hline
${\cal O}_7$ &$-\frac{i}{4}$ & & & &$-\frac{i}{4}$ &$-i$ & &
$\frac{in}{2(3n+1)}$ &$\frac{i(n+1)}{4(3n+1)}$ & &  & & & & & &\\
 \hline
${\cal O}_8$ & & & & & & &$i$ & & &  & & & & & & &\\
 \hline
${\cal O}_9$ &$-i$& & & & & & &$\frac{i}{2}$&$\frac{i}{2}$ &  & & & & & & &\\
 \hline
${\cal O}_{10}$ & & & & & & & &$-\frac{i}{2}$&$\frac{i}{2}$& & & & & &
& &\\
 \hline
${\cal O}_{11}$ & & & & &$-\frac{i}{4}$& &$-\frac{i}{2}$ & & & & &
$\frac{3i}{2}$&$2i$ &  &$-\frac{i}{2}$ & &\\
 \hline
${\cal O}_{12}$ &  & & & & & &  & & &$1$ & &$1$ & &$\frac{n+1}{3n+1}$ &$-\frac{n+1}{3n+1}$ & &\\
 \hline
${\cal O}_{13}$ &  & & & & & & & & & &$1$ & &$1$ & & & &\\
 \hline
${\cal O}_{14}$ &$\frac{i}{8}$& & & & & & & & & & &$-\frac{i}{4}$
&$-i$ &$\frac{in}{2(3n+1)}$ &$\frac{i(n+1)}{4(3n+1)}$ & &\\
 \hline
${\cal O}_{15}$ &$-\frac{i}{4}$& & & & & & & &$\frac{i}{2}$& & & & &
$\frac{i}{2}$ &$-\frac{3i}{2}$ & &$i$\\
 \hline
${\cal O}_{16}$ & & & & &  & & & & & & & & &$-\frac{i}{2}$ &$\frac{i}{2}$ & & $-\frac{i(n+1)}{3n+1}$\\
\hline
${\cal O}_{17}$ &$\frac{3}{16}$& & & & & & & &$-\frac{1}{8}$& & & & &
&$\frac{1}{2}$ & $-\frac{1}{2}$ &$-\frac{n+1}{4(3n+1)}$\\
\hline
\end{tabular}
\caption{\label{tab:1}
The coefficients in the
gauge variations of $CP$-even operators. Only non-zero entries
are shown.}
\end{table}
\end{landscape}

\begin{table}[h!]
\centering
\begin{tabular}{|c|c|c|c|c|c|c|c|c|}
\hline
&\begin{sideways} $\delta_L\big(w^aw^b\d_aH_b\DD_cH^c\big)$  \end{sideways}
&\begin{sideways} $\delta_L(w^aw^b\d_cH^c\DD_aH_b)$  \end{sideways}
&\begin{sideways} $\delta_L\big(w^aw^b\d_aH_b(\Gamma+\bar\Gamma)\big)$
\end{sideways}
&\begin{sideways} $\delta_L\big(iw^aw^b\DD_aH_b(\Gamma-\bar\Gamma)\big)$
\end{sideways}
&\begin{sideways} $\delta_L(\d_aH^a\DD_bH^b)$  \end{sideways}
&\begin{sideways} $\delta_L\big(\d_aH^a(\Gamma+\bar\Gamma)\big)$  \end{sideways}
&\begin{sideways}
  $\delta_L\big(i\DD_aH^a(\Gamma-\bar\Gamma)\big)$  \end{sideways}
&\begin{sideways} $\delta_L\big(i(\Gamma^2-\bar\Gamma^2)\big)$  \end{sideways}
\\

\hline
${\cal O}_1$ &$-\frac{i}{2}$ &$\frac{i}{2}$ & & & & &  &  \\
 \hline
${\cal O}_2$ &$\frac{i}{2}$ &$-\frac{i}{2}$ &$-\frac{i(n+1)}{3n+1}$
&$-\frac{i(n+1)}{3n+1}$  & & & & \\
 \hline
${\cal O}_5$ &$-\frac{1}{4}$ & & & & & &  &  \\
 \hline
${\cal O}_6$ & &$\frac{1}{4}$ & &$\frac{1}{2}$ & & & & \\
 \hline
${\cal O}_7$ &$\frac{3}{8}$ &$-\frac{1}{8}$ &$-\frac{2n+1}{2(3n+1)}$ &
$-\frac{n+1}{4(3n+1)}$ & & & & \\
 \hline
${\cal O}_9$ & & &$\frac{1}{2}$ &$-\frac{1}{2}$ & & & & \\
 \hline
${\cal O}_{10}$ & & &$\frac{1}{2}$ &$\frac{1}{2}$ & & & & \\
 \hline
${\cal O}_{11}$ & & & & &$-\frac{1}{4}$ & &$-\frac{1}{2}$ & \\
 \hline
${\cal O}_{12}$ & & & & & &$-\frac{i(n+1)}{3n+1}$ &$-\frac{i(n+1)}{3n+1}$ &\\
 \hline
${\cal O}_{14}$ & &$-\frac{1}{8}$ & & &$\frac{3}{8}$
&$-\frac{2n+1}{2(3n+1)}$&$-\frac{n+1}{4(3n+1)}$ &\\
 \hline
${\cal O}_{15}$ & & & &$-\frac{1}{2}$ & &$\frac{1}{2}$ &$\frac{3}{2}$ &\\
 \hline
${\cal O}_{16}$ &  & & & & &$\frac{1}{2}$ &$\frac{1}{2}$ &\\
 \hline
${\cal O}_{17}$ &  & & &$\frac{i}{8}$ & & &$-\frac{i}{2}$ &$\frac{i}{2}$ \\
 \hline
\end{tabular}
\caption{\label{tab:2}
Same as Table~\ref{tab:1}, but for the $CP$-odd operators.}
\end{table}

\section{The supercurrent multiplet}
\label{app:supercurrent}

Here we remind some general facts about the supercurrent
multiplet. Consider a supersymmetric theory coupled to
supergravity. The action can be written as
\[
S=\frac{1}{\vk^2} S_{SG} [g_{mn},\psi_{m\a}]+S_{\rm
  mat}[\Phi_i;g_{mn},\psi_{m\a}]
+\ldots\;,
\]
where $S_{SG}$ is the pure SUGRA action depending on the metric and
gravitino, $S_{\rm mat}$ is the action of the matter sector, and $\Phi_i$
collectively denotes the matter fields (both bosons and
fermions). We
assume that the auxiliary fields have been integrated out, so that we
are working only with the physical fields. Note that we have factored
out the (inverse of) the gravitational coupling $\vk^{-2}$ in front of
the SUGRA action. The matter action is of zeroth order in $\vk$. Dots
denote possible extra terms proportional to positive powers of
$\vk^2$ that appear upon integrating out the auxiliary fields.

Let us expand the action around Minkowski spacetime as
a power series in the metric perturbation $h_{mn}$ and
gravitino. Further, it is convenient to rescale the fields of the
gravity sector by introducing
\be
\label{canhpsi}
\hat h_{mn}=\vk^{-1} h_{mn}~,~~~~~\hat\psi_{m\a}=\vk^{-1}\psi_{m\a}\;.
\ee
One obtains,
\be
\label{Sexpand}
S= S^{(2)}_{SG}[\hat h_{mn},\hat \psi_{m\a}]
+S_{\rm mat}^{(0)}[\Phi_i]
+\vk \int d^4 x\bigg(\frac{1}{2}{\cal T}^{mn}\hat h_{mn}-
{\cal S}^{m\a}\hat\psi_{m\a}-\bar{\cal
  S}^m_{\dot\a}\bar{\hat\psi}^{\dot\a}_m\bigg)+\ldots\;,
\ee
where $S^{(2)}_{SG}$ is the quadratic part of the SUGRA action,
$S_{\rm mat}^{(0)}$ is the matter action in flat spacetime, ${\cal
  T}^{mn}$ and ${\cal S}^{m\a}$ are the energy-momentum ternsor (EMT)
and the supercurrent. Dots stand for terms that are of cubic
and higher order in $\hat h_{mn}$, $\hat\psi_{m\a}$ or are
proportional to $\vk^2$.

The action is invariant under local SUSY transformations which we also
expand as power series in $\hat h_{mn}$, $\hat\psi_{m\a}$
(but not in the matter fields). The parameter of these transformations
will be denoted by $\zeta_\a(x)$.
Then, up to terms quadratic in $\hat
h_{mn}$, $\hat\psi_{m\a}$, the transformations of the metric
perturbation and gravitino read,
\bseq
\label{dzetahpsi}
\begin{align}
\label{dzetah}
&\delta_\zeta\hat
h_{mn}=2i(\bar\zeta\bar\s_m\hat\psi_n+\bar\zeta\bar\s_n\hat\psi_m)+\text{h.c.}\;,\\
&\delta_\zeta\hat\psi_{m\a}=\frac{1}{\vk}\d_m\zeta_\a
+\frac{1}{2}(\s^{kn}\zeta)_\a \d_k\hat h_{nm}+\vk\,
\Xi_{m\a}[\Phi_i]+\ldots\;,
\label{dzetapsi}
\end{align}
\eseq
where the model-dependent spin-vector $\Xi_{m\a}$ is constructed of
matter fields and the omitted terms are of higher order in
$\vk$. Notice that the first of these equations coincides with
Eq.~(\ref{tildedGh}) which has the same form irrespectively of whether
$\zeta_\a$ is coordinate dependent or not. Whereas the gravitino
transformation is simply a combination of Eqs.~(\ref{standlocal}) and
(\ref{tildedGpsi}) with the identification $\varepsilon_\a=\zeta_\a$.
The matter field
transformations can also be written as an expansion in $\vk$,
\be
\label{dzetaPhi}
\delta_\zeta\Phi_i=\delta_\zeta^{(0)}\Phi_i+\vk\,
\delta_\zeta^{(1)}\Phi_i+\ldots\;,
\ee
where the first term is the transformation in flat spacetime and the
remaining terms describe corrections due to the supergravity
coupling. Then the invariance of the action implies,
\be
\label{dzetaS}
\begin{split}
&0=\int
d^4x
\bigg[\vk\big(4\epsilon^{klmn}\bar\Xi_k\bar\s_l\d_m\hat\psi_n+\text{h.c.}\big)
+\frac{\delta S^{(0)}_{\rm mat}}{\delta \Phi_i}
\big(\delta_\zeta^{(0)}\Phi_i+\vk\,\delta_\zeta^{(1)}\Phi_i\big)\\
&+\frac{\vk}{2}\delta_\zeta{\cal T}^{mn}\,\hat h_{mn}
+\frac{\vk}{2}{\cal T}^{mn}\,\delta_\zeta\hat h_{mn}
-\vk\big(\delta_\zeta{\cal S}^{m\a}\,\hat\psi_{m\a}
+{\cal
  S}^{m\a}\,\delta_\zeta\hat\psi_{m\a}+\text{h.c.}\big)\bigg]+\ldots\;,
\end{split}
\ee
where we have used the explicit form of the Rarita--Schwinger
Lagrangian for gravitino and took into account that the SUGRA action
itself is invariant in the absence of matter.
Let us consider the consequences of this equation order by order in
$\vk$.

At the zeroth order we have the variation of the flat-space matter
action. This is invariant under global SUSY with constant
$\zeta_\a$. Hence, its variation is proportional to the gradient of
$\zeta_\a$. The Noether theorem identifies the proportionality coefficient
with the supercurrent,
\[
\int d^4x\,\frac{\delta S_{\rm
    mat}^{(0)}}{\delta\Phi_i}\,\delta_\zeta^{(0)}\Phi_i
=\int d^4x\big({\cal S}^{m\a}\d_m\zeta_\a+\text{h.c.}\big)\;.
\]
We see that the gradient term in the transformation of gravitino
(\ref{dzetapsi}) precisely cancels this variation, as it should
be. This cancellation happens for arbitrary configuration
of the matter fields. On the other hand, we can restrict the fields
$\Phi_i$ on shell, i.e. consider only the configurations that satisfy
the flat-space equations of motion,
\be
\label{Phieom}
\frac{\delta S_{\rm
    mat}^{(0)}}{\delta\Phi_i}=0\;.
\ee
Then the variation of the matter action vanishes by itself and the
invariance of the remaining gravitino-matter coupling implies
$\d_m{\cal S}^{m\a}=0$, which is nothing but the on-shell conservation
of the SUSY current.

Consider now Eq.~(\ref{dzetaS}) at order
$O(\vk)$. In general, the matter fields'
transformations get corrected at this order
in a non-trivial model-dependent way.
Thus, inferring any off-shell statements is
problematic. However, the situation dramatically simplifies on shell, where
the variation of the matter action
vanishes due to Eqs.~(\ref{Phieom}).
Restricting also to constant $\zeta_\a$, so that the part
neglected in Eq.~(\ref{dzetaS}) cannot produce any $O(\vk)$
contribution, one is left with the terms containing variations of the
gravity fields, EMT and supercurrent. Substituting the explicit
formulas (\ref{dzetahpsi}) and equating the combinations in front of
$\hat h_{mn}$, $\hat\psi_{m\a}$ to zero one obtains the relations
(\ref{SCmult}) from the main text. Notice that the variations of the
matter fields inside ${\cal T}^{mn}$ and ${\cal S}^{m\a}$ are taken at
the
zeroth order corresponding to the flat-space global SUSY.

To sum up, we have shown that in any supersymmetric theory the EMT and
supercurrent describing the coupling of the theory to supergravity
belong to the same global SUSY multiplet with on-shell transformation
rules (\ref{SCmult}). The transformation of ${\cal T}^{mn}$ is
completely universal, whereas that of ${\cal S}^{m\a}$ depends on the
model through the spin-vector $\Xi^{m\a}$.

\section{Auxiliary field $\omega_\a$}
\label{app:omega}
In this Appendix we derive the on-shell value of the fermionic
auxiliary field $\omega_\a$ that enters into the SUSY transformation of
the aether. We treat the coefficient $C$ in the super-aether action
(\ref{superact}) as a small parameter and work up to terms of order
$O(\sqrt{C})$. In principle, one could use the same strategy as for
the bosonic sector: first derive the full off-shell fermionic
Lagrangian and then find from it the equations of motion of the
auxiliary fields. We find it simpler, however, to act in the reverse
order: first obtain the equations of motion in terms of superfields
and then project them on the appropriate components.

The superspace equations of motion are obtained by taking the
variation of the sum of the actions (\ref{NMact}) and (\ref{superact})
with respect to the superfields. One has to remember, however, that
the superfields $\Gamma$ and $V^a$ are constrained: $\Gamma$ is
linear, whereas $V^a$ satisfies the orthogonality and chirality
conditions
(\ref{Vnorm}), (\ref{Vchir}). We
implement these constraints using Lagrange multipliers. Thus, the
action is supplemented with the term,
\be
\label{Scons}
\begin{split}
S_{\rm constr}=\frac{1}{\vk^2}\int d^4x\, d^4\theta \Big[&
\bar\varLambda_1 \bar D^2\Gamma + \varLambda_1 D^2\bar\Gamma+
\bar\varLambda_2 w^a V_a+\varLambda_2 w^a \bar V_a\\
&+\bar\varLambda^{~a}_{3\dot\b}(\bar D^{\dot\b}V_a+w^c
\Phi^{\dot\b}_{~ca})
+\varLambda^{~\b}_{3a}(D_{\b}\bar V^a+w^c
\Phi_{\b c}^{\phantom{\b c}a})\Big]\;,
\end{split}
\ee
where the Lagrange multipliers $\varLambda_{1}$, $\varLambda_{2}$,
 $\varLambda_{3a}^{~\b}$ are general scalar and spin-vector
superfields.\footnote{An alternative way to enforce the orthogonality
  constraint (\ref{Vnorm}) is to use a chiral Lagrange multiplier as
  in Eq.~(\ref{soperpotential}).} Once this term is added, all
superfields in the action can be treated as unconstrained. Recalling
the expression (\ref{connvect}) for the connection in terms of the
SUGRA superfields and taking the variation with respect to
$\bar\Gamma$, $\bar V^a$ we obtain,
\bseq
\begin{align}
&i\frac{3n+1}{2n}\d_mH^m+\frac{3n+1}{2}\Delta_m H^m
+\frac{9n^2-1}{4n}\bar\Gamma+\frac{(3n+1)^2}{4n}\Gamma\notag\\
&\qquad+D^2\varLambda_1 +
w^c(\s_{ca})_{\b}^{~\,\gamma}D_\gamma\varLambda_3^{~a\b}=0\;,\\
&w_a\varLambda_2+D_\b\varLambda_{3a}^{~\b}+\frac{\sqrt{C}}{2}\hat V_a=0\;,
\end{align}
\eseq
where we have neglected terms of order $O(C)$ in the first equation
and in the second equation renormalized the aether superfield
canonically, $\hat V_a=\sqrt{C} V_a$. The neglected terms do not
affect the expression for $\omega_\a$ at order $O(\sqrt{C})$. We now
take the covariant derivative $D_\a$ of these two equations and
restrict them to the origin of spinor coordinates $\theta$,
$\bar\theta$. Using the definition of the component fields
(\ref{Hcomp}), (\ref{compGamma}), (\ref{Vcomp}) we find,
\bseq
\begin{align}
&-\frac{3n+1}{2n}\,\omega_{\a}
+\frac{1}{2}w^c(\s_{ca})_{\b\a}D^2\varLambda_3^{~a\b}\big|=0\;,\\
&w_a D_\a\varLambda_2\big| +\frac{1}{2}D^2\varLambda_{3a\a}\big|
+\frac{\sqrt{C}}{2}\hat\eta_{a\a}=0\;,
\end{align}
\eseq
where we again neglected the terms of higher order in $C$ and used the
gauge conditions (\ref{WZ}), (\ref{superconffix}). Combining these
equations to eliminate the Lagrange multipliers we arrive at the expression
(\ref{omega}) from the main text.

\section{Velocities of elementary excitations}
\label{app:disph1}

\subsection{Bosonic modes}
\label{app:vbos}

Here we analyze in detail the sector of modes described
by the Lagrangian (\ref{Ltot}). We work in the frame where the aether
VEV has vanishing spatial components, see Eq.~(\ref{wtimel}). In this frame
the time-components of aether perturbations vanish due to the
constraint (\ref{Vnorm}), $\hat v^{R,I}_0=0$.

Transverse traceless
(helicity $\pm 2$) modes are contained only in the metric. Thus, we
insert the Ansatz,
\[
h_{ij}=h^{tt}_{ij}~,~~~~~~\d_i h_{ij}^{tt}=h_{ii}^{tt}=0~,~~~~~i,j=1,2,3,
\]
with all other fields vanishing. The Lagrangian becomes
\be
\label{Lh2}
{\cal L}_{h=2}=\frac{1}{2\vk^2}\bigg[-\frac{1-C}{4}h_{ij}^{tt}
{\ddot h}_{ij}^{tt}+\frac{1}{4}h_{ij}^{tt}\Delta
{h}_{ij}^{tt}\bigg]\;,
\ee
where we use $\DD$ to denote the
spatial Laplacian,
$\DD=\d_i\d_i$; as we are not going to use the operator
(\ref{deltadef}) in this Appendix, this should not lead to
confusion. Decomposing the field into plane waves with momentum ${\bf
  p}$ and energy $E$ we find the dispersion relation
\be
\label{dispgrav}
E^2=\frac{p^2}{1-C}\approx (1+C)\,p^2\;,
\ee
which corresponds to the propagation velocity (\ref{sgrav}).

We now turn to modes with helicities $\pm 1$.
The metric perturbations
are taken in the form
\[
h_{00}=0~,~~~~h_{0i}=n_i~,~~~~h_{ij}=\d_i\xi_j+\d_j\xi_i\;,
\]
with all vectors being transverse,
\[
\d_in_i=\d_i\xi_i=\d_i\hat v^R_i=\d_i\hat v^I_i=0\;.
\]
The Lagrangian (\ref{Ltot}) in this sector reads,
\be
\label{Lh1}
\begin{split}
{\cal L}_{h=1}=&\frac{1}{2\vk^2}\bigg[
-\frac{1}{2}n_i \DD n_i
+\frac{1}{2}\xi_i\DD\ddot\xi_i-\dot n_i\DD\xi_i
-\hat v^R_i\ddot{\hat v}^R_i
+\hat v^R_i\DD{\hat v}^R_i\\
&-\bigg(1-\frac{C}{2}\bigg)\hat v^I_i\ddot{\hat v}^I_i
+\hat v^I_i\DD{\hat v}^I_i
-\sqrt{C}\hat v_i^R\ddot n_i
+\sqrt{C}\hat v_i^R\DD\dot \xi_i
+C\epsilon_{ijk}\,\hat v_i^R \d_j\dot{\hat v}^I_k \bigg]\;,
\end{split}
\ee
where $\epsilon_{ijk}$ is the 3-dimensional antisymmetric
tensor, $\epsilon_{123}=1$.
In deriving this expression we kept only the
leading-order terms in the gravitational part of the Lagrangian: the
omitted corrections affect the dynamics of helicity-1 modes only at order
$O(C^{3/2})$ or higher. Note a peculiar mixing term between
the real and imaginary parts of the aether perturbations.

Varying (\ref{Lh1}) with respect to $\xi_i$ and setting the gauge
$\xi_i=0$ afterwards one finds,
\be
\label{nvR}
n_i=-\sqrt{C}\,\hat v_i^R
\ee
up to corrections of order\footnote{The variation of (\ref{Lh1}) with
  respect to $n_i$ yields the equation
\be
\label{nvR2}
\DD n_i+\sqrt{C}\, \ddot{\hat v}_i^R=0\;.
\ee
By combining this with (\ref{nvR}) one could naively conclude that the
velocity of the excitations described by $\hat v_i^R$ is equal to
1. This is true only at the zeroth order in $C$: equations
(\ref{nvR}), (\ref{nvR2}) are valid only up to $O(C^{3/2})$
corrections and hence do not allow to capture the $O(C)$ terms in the velocity,
which we are interested in.
}
$O(C^{3/2})$. Substituting this into the equations for the aether
perturbations we obtain,
\bseq
\label{vIR}
\begin{align}
\label{vIR1}
&-\bigg(1-\frac{C}{2}\bigg)\ddot{\hat v}^R_i+\DD\hat v^R_i
+\frac{C}{2}\epsilon_{ijk}\d_j\dot{\hat v}^I_k=0\;,\\
\label{vIR2}
&-\bigg(1-\frac{C}{2}\bigg)\ddot{\hat v}^I_i+\DD\hat v^I_i
-\frac{C}{2}\epsilon_{ijk}\d_j\dot{\hat v}^R_k=0\;.
\end{align}
\eseq
To solve this system, we take $\hat v^{R,I}_i$ as a sum of circularly
polarized plane waves with energy $E$ and momentum $\bm{p}$,
\[
\hat v_i^{R,I}=\big(e_i^{(+)}f_+^{R,I}+e_i^{(-)}f_-^{R,I}\big)\;
\e^{-iEt+i\bm{px}}\;,
\]
where
\[
e_i^{(\pm)}=e_i^{(1)}\pm i \,e^{(2)}_i\;,
\]
and the unit vectors $\bm{e}^{(1)}$, $\bm{e}^{(2)}$ form together with
$\bm{e}^{(3)}\equiv\bm{p}/p$ a right-handed triad. Substituting these expressions into
Eqs.~(\ref{vIR}) and diagonalizing the resulting eigenvalue matrix we
find the dispersion relations for the modes:
\begin{itemize}
\item[i)]
modes with $f_\pm^I=\pm if_\pm^R$~~~$\Longrightarrow$~~~ $E^2=p^2$
\item[ii)]
modes with $f_\pm^I=\mp if_\pm^R$~~~$\Longrightarrow$~~~ $E^2=(1+C)\, p^2$
\end{itemize}
In the latter case the dispersion relation coincides with that of
gravitons, see Eq.~(\ref{dispgrav}),
which identifies the corresponding modes as members of the
gravitational supermultiplet.
Note that these modes are an essential mixture of
real and imaginary aether components with the admixture of metric
perturbations, see Eq.~(\ref{nvR}).

Let us comment on the consequences of SUSY breaking. As
discussed in \cite{Pujolas:2011sk}, it leads to the generation of mass
for the imaginary part of the aether $\hat v^I_i$. Then the dispersion
relation for the remaining component $\hat v^R_i$ is obtained from
(\ref{vIR1}) by simply dropping off the last term, which yields
$E^2=(1+C/2)\,p^2$. We conclude that the SUSY breaking modifies the
velocity of the helicity 1 modes, so that its deviation from unity is
twice smaller than that for gravitons. This coincides with the result
in the Einstein-aether model \cite{Jacobson:2004ts} for the choice of
parameters (\ref{SGc}).

Finally, we consider the helicity 0 sector. Here the Ansatz reads,
\[
h_{00}={\cal N}~,~~~~h_{0i}=\d_i{\cal
  B}~,~~~~h_{ij}=\delta_{ij}\,\varphi+\d_i\d_j{\cal E}~,~~~~
\hat v^{R,I}=\d_i\upsilon^{R,I}\;.
\]
Substitution into the Lagrangian yields,
\be
\label{Lh0}
\begin{split}
{\cal L}_{h=0}=&\frac{1}{2\vk^2}\bigg[\frac{3}{2}\varphi\ddot\varphi
-\frac{1}{2}\varphi\Delta\varphi+{\cal N}\Delta\varphi
-2\varphi\Delta\dot{\cal B}+\varphi\Delta\ddot{\cal E}
+\upsilon^R\Delta\ddot\upsilon^R-\upsilon^R\Delta^2\upsilon^R\\
&-\sqrt{C}\upsilon^R\Delta\dot\varphi
-\sqrt{C}\upsilon^R\Delta\dot{\cal N}
+\sqrt{C}\upsilon^R\Delta\ddot{\cal B}
+\sqrt{C}\upsilon^R\Delta^2 {\cal B}
-\sqrt{C}\upsilon^R\Delta^2\dot{\cal E}\\
&+\bigg(1-\frac{C}{2}\bigg)\upsilon^I\Delta\ddot\upsilon^I
-\bigg(1-\frac{C}{2}\bigg)\upsilon^I\Delta^2\upsilon^I
\bigg]\;,
\end{split}
\ee
where we again omitted terms of order $O(C)$ in the purely gravitational
part. We see right away that the mode $\upsilon^I$ decouples and has
unit velocity. The mode associated with the real part of the aether
requires a bit more work.
Varying with respect to ${\cal N}$ and $\varphi$ and then
setting ${\cal N}={\cal B}=0$ by a gauge fixing we obtain the
constraints,
\[
\Delta\varphi=-\sqrt{C}\Delta\dot\upsilon^R\;,\qquad
\Delta\ddot{\cal
  E}=-3\ddot\varphi+\Delta\varphi-\sqrt{C}\Delta\dot\upsilon^R\notag\;.
\]
Substituting the above expressions into the equation
obtained by variation with respect to $\upsilon^R$,
\[
2\Delta\ddot\upsilon^R-2\Delta^2\upsilon^R
-\sqrt{C}\Delta\dot\varphi-\sqrt{C}\Delta^2\dot{\cal E}=0\;,
\]
we observe that $\upsilon^R$ satisfies the relativistic wave
equation. Thus, this mode also has unit velocity.
One can check that variation with respect to ${\cal B}$ and ${\cal E}$
does not give any
new relations.

\subsection{Fermionic modes}
\label{app:vferm}

In the fermionic sector we find it more convenient to work directly
with the equations of motion. From the Lagrangian (\ref{Sferm}) we
have,
\bseq
\label{fermeq1}
\begin{align}
4\epsilon^{abcd} \bar\s_b\d_c\psi_d-\sqrt{C} w^a\d_b\bar{\hat\eta}^b
+\sqrt{C}w^b\d_b\bar{\hat\eta}^a=0\;,\\
-i\s^b\d_b\bar{\hat\eta}_a+4\sqrt{C} w^b\d_a\psi_b-4\sqrt{C} w^b\d_b\psi_a=0\;.
\end{align}
\eseq
We again perform the $(3+1)$ split of all quantities into temporal and
spatial components. Upon reduction to the spatial rotations, the
dotted and undotted spinor indices can be identified because the
fundamental representation of $SU(2)$ is equivalent to its complex
conjugate. More precisely, a spinor $\upchi_\a$ with lower index and
its complex conjugate with upper index $\bar\upchi^{\dot\a}$ transform
in the same way and can be treated as just two-component columns. We
impose the gauge $\psi_{0\a}=0$ and recall that $\bar{\hat\eta}_0^{\dot\a}=0$ due
to the orthogonality condition (\ref{Vnorm}). Further, we use the
explicit form of the $\s$-matrices \cite{WB},
$\s_0=\bar\s_0=\mathbbm{1}$,
$\bar\s_i=-\s_i$,
where $\s_i$, $i=1,2,3$, are the usual Pauli matrices. Then Eqs.~(\ref{fermeq1})
imply,
\bseq
\label{fermeq2}
\begin{align}
\label{fermeq21}
&-4\epsilon_{ijk}\s_i\d_j\psi_k-\sqrt{C}\d_i\bar{\hat\eta}_i=0\;,\\
&-4\epsilon_{ijk}\s_j\dot\psi_k-4\epsilon_{ijk}\d_j\psi_k
+\sqrt{C}\dot{\bar{\hat\eta}}_i=0\;,\\
&i\dot{\bar{\hat\eta}}_i-i\s_j\d_j\bar{\hat\eta}_i-4\sqrt{C}\dot\psi_i=0\;,
\end{align}
\eseq
where all tensor indices run from 1 to 3.
Note that the first equation (\ref{fermeq21}) does
not contain time derivatives and thus represents a constraint
reflecting the gauge invariance of the gravitino field.

Next step is to perform decomposition into plane waves. In addition,
we decompose the fields in the basis of unit vectors $\bm{e}^{(r)}$,
$r=1,2,3$, that form a right-handed triad, $\bm{e}^{(3)}$ being aligned
with the momentum (cf. Sec.~\ref{app:vbos}). Thus we write,
\be
\label{fermdecomp}
\psi_i=\Big(\sum_r F_{(r)}\, e_i^{(r)}\Big) \e^{-iEt+i\bm{px}}~,~~~~
\bar{\hat\eta}_i=\Big(\sum_r G_{(r)}\, e_i^{(r)}\Big) \e^{-iEt+i\bm{px}}\;,
\ee
where $F_{(r)}$, $G_{(r)}$ are spinor coefficients. Substituting into
Eqs.~(\ref{fermeq2}) we obtain,
\bseq
\label{fermeq3}
\begin{align}
&4 (\bm{\s e}^{(1)}) F_{(2)}-4(\bm{\s e}^{(2)}) F_{(1)}
-\sqrt{C}\, G_{(3)}=0\;,\\
&4\big[-E (\bm{\s e}^{(3)}) +p\big] F_{(2)} +4 E
(\bm{\s e}^{(2)}) F_{(3)}
-\sqrt{C}E\,G_{(1)}=0\;,\\
&4\big[E (\bm{\s e}^{(3)}) -p\big] F_{(1)} -4 E
(\bm{\s e}^{(1)}) F_{(3)}
-\sqrt{C}E\,G_{(2)}=0\;,\\
&\big[E+(\bm{\s p})\big]G_{(r)}+4i\sqrt{C}E\,F_{(r)}=0~,~~~~~~r=1,2,3\;.
\end{align}
\eseq
The system is simplified by introducing the linear combinations,
\be
\label{FGpm}
F_{(\pm)}=(\bm{\s e}^{(2)})F_{(1)}\pm
(\bm{\s e}^{(1)})F_{(2)}~,~~~~~~
G_{(\pm)}=(\bm{\s e}^{(1)})G_{(1)}\mp
(\bm{\s e}^{(2)})G_{(2)}\;.
\ee
Then the $(+)$ and $(-)$ modes decouple.
In the $(+)$ sector we obtain,
\bseq
\label{fermeq+}
\begin{align}
\label{fermeq+1}
&4\big[E(\bm{\s e}^{(3)})+p\big] F_{(+)}-\sqrt{C}E\,G_{(+)}=0\;,\\
\label{fermeq+2}
&\big[E(\bm{\s e}^{(3)})-p\big] G_{(+)}-4\sqrt{C}E\,F_{(+)}=0\;.
\end{align}
\eseq
Mupliplying (\ref{fermeq+1}) by the operator
$\big[E(\bm{\s e}^{(3)})-p\big]$ and substituting $G_{(+)}$ from
(\ref{fermeq+2}) we arrive at a linear equation for $F_{(+)}$ which implies
the dispersion relation $E^2\approx (1+C)\,p^2$ coinciding with
that of gravitons. One can show that the $(+)$ modes have helicities
$\pm 3/2$. For a given
momentum there are two linearly independent modes with positive
energy and two modes with negative energy. Thus, in total we find four
helicity $\pm 3/2$ states that belong to the graviton multiplet, as
discussed in Sec.~\ref{sec:spectrum}.

The equations in the $(-)$ sector read,
\bseq
\label{fermeq-}
\begin{align}
&-4F_{(-)}-\sqrt{C}\,G_{(3)}=0\;,\\
&-4\big[E(\bm{\s e}^{(3)})+p\big]F_{(-)}+8iE(\bm{\s e}^{(3)})F_{(3)}
-\sqrt{C}E\,G_{(-)}=0\;,\\
&\big[E(\bm{\s e}^{(3)})-p\big]G_{(-)}-4\sqrt{C}E\,F_{(-)}=0\;,\\
&\big[E(\bm{\s e}^{(3)})+p\big]G_{(3)}+4i\sqrt{C}E (\bm{\s e}^{(3)}) F_{(3)}=0\;.
\end{align}
\eseq
Expressing $F_{(-)}$ and $(\bm{\s e}^{(3)})F_{(3)}$ from the first
two equations and substituting the result back into the third and
fourth ones we arrive at,
\begin{align}
&\big[E(\bm{\s e}^{(3)})-p\big]G_{(-)}+CE\,G_{(3)}=0\;,\notag\\
&(2-C) \big[E(\bm{\s e}^{(3)})+p\big]G_{(3)}+CE\,G_{(-)}=0\;.\notag
\end{align}
We see that the mixing between $G_{(-)}$ and $G_{(3)}$ is of order
$O(C)$. It contributes to the dispersion relations only at higher
orders and can be neglected. Then we
obtain two decoupled equations for the modes $G_{(-)}$ and $G_{(3)}$
leading to relativistic dispersion relations $E^2=p^2$. There are two
linearly independent modes with positive energy and two modes with
negative energy. In total, they comprise four helicity $\pm 1/2$
states that complement the fermionic sector of the theory.

\end{document}